\documentclass[11pt,a4paper]{article}
\usepackage{jcappub,natbib}
\usepackage{braket}
\input{colordvi.tex}

\title{CMB power spectra induced by primordial cross-bispectra between metric perturbations and vector fields
}
\author[a]{Maresuke Shiraishi,}
\author[a]{Shohei Saga}
\author[b]{and Shuichiro Yokoyama}
\affiliation[a]{Department of Physics and Astrophysics, Nagoya University, 
Nagoya, Aichi, 464-8602, Japan}
\affiliation[b]{Institute for Cosmic Ray Research, University of Tokyo, 5-1-5 Kashiwa-no-Ha, Kashiwa, Chiba, 277-8582, Japan}
\emailAdd{mare@nagoya-u.jp}
\emailAdd{saga.shohei@nagoya-u.jp}
\emailAdd{shu@icrr.u-tokyo.ac.jp}
\abstract{We study temperature and polarization anisotropies of the cosmic microwave background (CMB) radiation sourced from primordial cross-bispectra between metric perturbations and vector fields, which are generated from the inflation model where an inflaton and a vector field are coupled. In case the vector field survives after the reheating, both the primordial scalar and tensor fluctuations can be enhanced by the anisotropic stress composed of the vector fields during radiation dominated era. We show that through this enhancement the primordial cross-bispectra generate not only CMB bispectra but also CMB power spectra. In general, we can expect such cross-bispectra produce the non-trivial mode-coupling signals between the scalar and tensor fluctuations. However, we explicitly show that such mode-coupling signals do not appear in CMB power spectra. Through the numerical analysis of the CMB scalar-mode power spectra, we find that although signals from these cross-bispectra are smaller than primary non-electromagnetic ones, these have some characteristic features such as negative auto-correlations of the temperature and polarization modes, respectively. On the other hand, signals from tensor modes are almost comparable to primary non-electromagnetic ones and hence the shape of observed $B$-mode spectrum may deviate from the prediction in the non-electromagnetic case. The above imprints may help us to judge the existence of the coupling between the scalar and vector fields in the early Universe.} 
\arxivnumber{1209.3384}
\begin{document}
\maketitle
\flushbottom
\allowdisplaybreaks[4]


\def\up{\;\raise1.0pt\hbox{$'$}\hskip-6pt\partial\;}
\def\down{\;\overline{\raise1.0pt\hbox{$'$}\hskip-6pt\partial\;}}
\section{Introduction} 

Owing to accurate observations and data treatments, it has turned out that galaxies and clusters of galaxies hold micro-gauss magnetic fields (e.g., \cite{Kronberg:2007dy, Bernet:2008qp, Wolfe:2008nk, Fletcher:2010wt}). Furthermore, studies on some astrophysical processes suggest that the strength of magnetic field in the inter-galactic medium is larger than ${\cal O}(10^{- 15} - 10^{-20})$ gauss \cite{Neronov:1900zz, Dolag:2010ni, Tavecchio:2010mk, Takahashi:2011ac}. What is the origin of these large-scale magnetic fields? It may be generated via an interaction between inflatons and some sort of vector fields, which breaks the conformal invariance, in the inflationary Universe 
(e.g., \cite{Grasso:2000wj, Bamba:2003av,  Bamba:2006ga, Martin:2007ue, Demozzi:2009fu}) \footnote{Other than the inflationary origins, diverse generation and amplification mechanisms of the magnetic field  have been proposed (e.g., \cite{Widrow:2002ud, Giovannini:2003yn, Ichiki:2006cd, Fenu:2010kh}).}
and such kind of model can generate nano-Gauss magnetic fields at the present Universe \cite{Martin:2007ue}, which is comparable to upper limits from the power spectra \cite{Durrer:1998ya, Mack:2001gc, Lewis:2004ef, Yamazaki:2008gr, Paoletti:2008ck, Finelli:2008xh, Shaw:2009nf, Yamazaki:2010nf, Shaw:2010ea, Paoletti:2010rx, Yadav:2012uz, Paoletti:2012bb}, bispectra \cite{Seshadri:2009sy, Caprini:2009vk, Trivedi:2010gi, Cai:2010uw, Shiraishi:2010yk, Shiraishi:2011fi, Shiraishi:2011dh, Shiraishi:2012rm} and trispectra \cite{Trivedi:2011vt} of CMB anisotropies. 
However, the models with such coupling might suffer from some problems.
One is that the energy density of the vector field spoils the inflationary
background dynamics, so-called, backreaction problem,
and another is that due to the time-dependence of the coupling 
the Universe enters strong coupling regime
at the early stage of inflation, so-called strong coupling problem
(e.g., see ref. \cite{Demozzi:2009fu}).
Hence, it has been commonly understood that
constructing a realistic model of generating primordial
magnetic fields during inflation is quite difficult.
There are also several works about universal bounds for the inflationary model
from the current observations of the large scale magnetic fields \cite{Demozzi:2012wh, Suyama:2012wh, Fujita:2012rb}. 

Recently, such vector field and coupling have been also widely discussed beyond the context of magnetogenesis (e.g., \cite{Gumrukcuoglu:2007bx, Kanno:2010nr, Watanabe:2010fh, Sorbo:2011rz, Barnaby:2010vf, Barnaby:2011vw, Barnaby:2012tk}). Because, the interaction between the inflaton and the vector field can induce the cosmological three-point cross-correlation (cross-bispectrum) between the fluctuations of the scalar field, metric perturbations and electromagnetic fields and there must be some cosmological imprints of it. 

In ref.~\cite{Caldwell:2011ra,Motta:2012rn}, 
although in the context of the generation of the magnetic field, the authors computed such a cross-bispectrum composed of one scalar field and two magnetic fields, and analyzed its shape dependence in the similar manner as the cases of non-Gaussianities without magnetic fields \cite{Babich:2004gb, Komatsu:2010hc, Komatsu:2010fb}. In refs.~\cite{Jain:2012ga, Jain:2012vm}, the authors developed the magnetic consistency relation similar to the consistency relation in the local-type non-Gaussianity. They suggested that this cross-bispectrum may be accessible through the cosmic microwave background (CMB) observations and a combined survey of large scale structure and Faraday rotation \cite{Stasyszyn:2010kp}. Therefore closely estimating CMB signals generated from the cross-bispectrum will be an interesting and important work for studying the early Universe. Furthermore, although refs.~\cite{Caldwell:2011ra, Motta:2012rn} have not discussed, the cross-bispectra involving tensor perturbation or the electric field can also be generated and hence may create characteristic CMB fluctuations. In refs.~\cite{Barnaby:2012tk, Barnaby:2012xt},
the authors investigate the primordial fluctuations sourced from the vector field through the coupling
during inflation and estimate the non-Gaussianity of the primordial curvature fluctuations.

This paper examines impacts of such electromagnetic-scalar and electromagnetic-tensor bispectra on CMB anisotropies. Electromagnetic fields, which are created during inflation, induce the anisotropic stress fluctuation (depending quadratically on electromagnetic fields) and affect the subsequent evolution of the CMB fluctuation of scalar, vector and tensor modes even during the radiation dominated era \cite{Mack:2001gc, Lewis:2004ef, Shaw:2009nf}. This implies that the electromagnetic-scalar and electromagnetic-tensor bispectra become sources of not only CMB bispectra but also CMB power spectra. Hence, first, we construct complete formulae for these cross-bispectra and the power spectra of temperature and polarization anisotropies. In such case,
since the scalar and tensor modes sourced from such anisotropic stress fluctuations have same origins, it could be readily imagined that the scalar-tensor mode coupling appears.
However, we explicitly show that the CMB power spectra do not have such mode-coupling components. 
Then, we find that CMB power spectra from the electromagnetic-scalar and electromagnetic-tensor bispectra have characteristic features. In the power spectrum of the scalar modes, although these are subdominant contributions compared with the primary non-electromagnetic signals \cite{Komatsu:2010fb}, the auto-correlations of the temperature (intensity) and $E$-mode polarization have negative values, respectively. CMB power spectra induced from the tensor modes have almost the same amplitude of primary ones. In this sense, the effects on the $B$-mode spectrum might be quite interesting. 

This paper is organized as follows. In the next section, we summarize some behaviors of the vector field and metric perturbations in the context of slow-roll inflation. In section~\ref{sec:bispectrum}, we present the formulae for the electromagnetic-scalar and electromagnetic-tensor bispectra. In section~\ref{sec:cmb}, we compute and analyze the CMB signals generated from these cross-bispectra. The final section is devoted to the summary and discussion of this paper. 

Throughout this paper, we obey the definition of the Fourier transformation as 
\begin{eqnarray}
f({\bf x}) \equiv \int \frac{d^3 {\bf k}}{(2 \pi)^3} \tilde{f}({\bf k}) e^{i {\bf k} \cdot {\bf x}}~,
\end{eqnarray}
and the rule that the subscripts and superscripts of the Greek characters and alphabets run from 0 to 3 and from 1 to 3, respectively.

\section{Nature of inflation coupled with the vector field} \label{sec:inflation}

In this section, let us discuss some background and perturbative quantities in a simple inflationary model with the coupling between a scalar field (inflaton) and a vector field. The action is given by 
\begin{eqnarray}
S &=& \int d^4x \sqrt{-g}\left[ -\frac{M_{\rm pl}^{2}}{16\pi}R-\frac{1}{2}g^{\mu \nu}\partial_{\mu}\varphi\partial_{\nu}\varphi -V(\varphi) \right] + S_V ~, \\
S_V &=& - \int d^{4}x 
\frac{1}{4}\sqrt{-g} g^{\mu \lambda}g^{\nu \sigma}W(\varphi)F_{\mu \nu}F_{\lambda \sigma} ~, \label{eq:EM_action}
\end{eqnarray}
where $\varphi$ is the inflaton, $F_{\mu \nu} \equiv \partial_\mu A_\nu - \partial_\nu A_\mu$ with $A_\mu$ being the vector field, and $M_{\rm pl} \equiv 1/\sqrt{G}$ is the Planck mass. For the dynamics of the vector field, only the time-dependence of the coupling function is important. Here, we simply impose the power-law type as 
\begin{eqnarray}
W(\eta) = W_I \left( \frac{a(\eta)}{a_I} \right)^{2n}~, 
\end{eqnarray}
where $a(\eta)$ is the scale factor with $\eta$ being the conformal time, and the subscript $I$ means the value at the end of inflation, $\eta=\eta_I$. This power-law shape is often realized by an assumption of a dilaton-like coupling \cite{Kanno:2009ei, Motta:2012rn}. Due to the time dependence of this coupling, the conformal invariance of the action (\ref{eq:EM_action}) is broken and hence the vector field can be survive even in the inflationary expansion. In addition, let us consider the case that after the end of inflation, $W = W_I$ is kept and the conformal invariance is restored. 
This seems to be natural assumption, since the coupling function $W$ is, here, dependent on the inflaton and in general the inflaton falls into a stable state after the end of inflation.
Owing to this condition, we do not need to consider the additional growth of the vector field at late stages, which is disfavored by observations. In the literature, to identify such the vector field with the observed magnetic field, the coupling function at the end of inflation $W_I$ should be taken
to be  unity \cite{Grasso:2000wj, Bamba:2003av,  Bamba:2006ga, Martin:2007ue, Demozzi:2009fu}. 

The evolution of the vector field strongly depends on the power-law index of the coupling function, $n$. In order not to spoil inflation due to the backreaction of the energy density of the vector field,
we have a bound on the spectral tilt of the running coupling: $|n| \lesssim 2.1$ (e.g., see \cite{Martin:2007ue, Demozzi:2009fu,Motta:2012rn}). 

This type of magnetogenesis model has another problem.
In case where $n > 0$ and $W_I$ is taken to be unity,
the running coupling constant, $W(\eta)$, is so small at the beginning of inflation and it means that the gauge coupling at the beginning of inflation is quite large, that is, inflation starts at the strong coupling regime. 
In such regime, we can not treat the perturbation of the vector field as a free field any more.
However, refs.~\cite{Caldwell:2011ra, Motta:2012rn} pointed out the possibility to avoid such a strong coupling problem by the action arising from the UV completion and including the violation of the gauge invariance as described in theories with extra dimensions. On the other hand, in ref.~\cite{Barnaby:2012tk}, the authors mentioned another possibility that the vector field localizes in a hidden sector and generates real electromagnetic field through any hidden coupling. In the following discussions, although the solution of the strong coupling remains unclear, we shall admit the above treatments and continue to discuss about cosmological imprints of the vector field.


In this paper, we assume the single field slow-roll inflation model where the slow-roll parameters, $\epsilon \equiv -\dot{H}/H^2$ and $\delta \equiv \ddot{\varphi}/(H \dot{\varphi})$, are almost constant and much less than unity. Here $H$ is the Hubble parameter and $~\dot{}~ \equiv \partial / \partial t$ denotes the derivative with respect to the physical time. Furthermore, following refs.~\cite{Caldwell:2011ra, Motta:2012rn}, we obey the isotropic background metric and perturbatively evaluate impacts of the vector field. One can see comprehensive analyses of the inflationary perturbations on the anisotropic background in e.g., refs.~\cite{Gumrukcuoglu:2007bx, Gumrukcuoglu:2010yc, Watanabe:2010fh}. Then, we have the relations between background quantities as
\begin{eqnarray}
\eta = - \frac{1}{aH} \frac{1}{1-\epsilon} ~, \ \
\frac{a}{a_I} = \left( \frac{\eta}{\eta_I} \right)^{1/(\epsilon-1)}~.
\end{eqnarray}

\subsection{Evolutions of the fluctuations}

Let us evaluate the evolution of curvature and tensor perturbations, and the fluctuation of the vector field, following ref. \cite{Motta:2012rn}. 
Using the Arnowitt-Deser-Misner formalism, each component of the scalar- and tensor-mode metric up to the first order is given by
\footnote{For decaying nature, we neglect the vector-mode metric perturbation.}
\begin{eqnarray}
 g_{00} = -N^2 + g_{ij} N^i N^j ~, \ \ 
g_{0i} = g_{ij} N^j ~, \ \ 
g_{ij} = a^2 [ (1 + 2 {\cal R}) \delta_{ij} + h_{ij} ] ~, \label{eq:metric}
\end{eqnarray}
 with 
\begin{eqnarray}
N = 1 + \frac{\dot{\cal R}}{H} ~, \ \ 
N^i = -\frac{\partial_i {\cal R}}{a H} + \epsilon \partial_i \nabla^{-2} \dot{\cal R}~. 
\end{eqnarray}
Here, we obey the gauge, $\delta \varphi = 0$, and the transverse-traceless condition, $\partial^i h_{ij} = h^i_{~i} = 0$. 
These metric perturbations and the vector field are quantized as
\begin{eqnarray}
\begin{split}
\mathcal{R}({\bf x},\eta) 
&= \int \frac{d^{3} {\bf k}}{(2\pi)^{3}}\left[ \mathcal{R}_{k}(\eta)e^{i{\bf k}\cdot {\bf x}}\beta({\bf k}, 0)+h.c. \right] ~, \\ 
A_{i}({\bf x},\eta) 
&= \int \frac{d^{3} {\bf k}}{(2\pi)^{3}} 
\sum_{\lambda = \pm 1} \left[ v_{k}(\eta)e^{i{\bf k}\cdot {\bf x}}O^{(\lambda)}_{i}(\hat{\bf k})\beta({\bf k},\lambda) + h.c. \right] ~, \\
h_{ij}({\bf x},\eta) 
&= \int\frac{d^{3}{\bf k}}{(2\pi)^{3}}\sum_{\lambda = \pm 2}
\left[ h_{k}(\eta) e^{i{\bf k}\cdot {\bf x}} O^{(\lambda)}_{ij}(\hat{\bf k}) \beta({\bf k}, \lambda) + h.c. 
\right] ~,
\end{split}
\end{eqnarray}
where $\beta({\bf k}, \lambda)$ and $\beta^\dagger({\bf k}, \lambda)$ are
respectively the annihilation and creation operators of ${\cal R}~(\lambda = 0)$, $A_i~(\lambda = \pm 1)$ and $h_{ij}~(\lambda = \pm 2)$, and satisfy $\beta({\bf k}, \lambda) \Ket{ 0} = 0$ and $[\beta({\bf k}, \lambda), \beta^\dagger({\bf k'}, \lambda') ] = (2\pi)^3 \delta_{\lambda, \lambda'} \delta({\bf k} - {\bf k'})$. In terms of the vector field, we have adopted the Coulomb gauge $\partial^i A_i = 0$ and $A_0 = 0$. 
The explicit forms of the
divergenceless vector, $O^{(\pm 1)}_{i}$ and transverse-traceless tensor, $O^{(\pm 2)}_{ij}$, are shown in appendix~\ref{appen:polarization}.
 Then, 
 as the mode function of each perturbation, we have \cite{Motta:2012rn, Stewart:1993bc}
\begin{eqnarray}
\begin{split}
\mathcal{R}_{k}(\eta) &= \mathcal{R}_{k*}u_{\nu}(-k\eta) ~, \\
v_{k}(\eta) &= v_{k*}u_{\alpha}(-k\eta) ~, \\
h_{k}(\eta) &= h_{k*}u_{\mu}(-k\eta) ~, 
\end{split}
\end{eqnarray}
where we have defined the time-independent and time-dependent functions, respectively, as  
\begin{eqnarray}
\mathcal{R}_{k*} &=& ie^{i\pi\nu/2+i\pi/4}(1-\epsilon)2^{\nu}\Gamma(\nu)\frac{H_*}{\sqrt{\epsilon}M_{\rm pl}}\frac{(-k\eta_*)^{3/2-\nu}}{k^{3/2}} ~, \label{eq:scal_mode_coeff} \\ 
v_{k*} &=& -i e^{i\pi\alpha /2 +i\pi /4}
 \frac{2^{\alpha -1}\Gamma(\alpha)}{\pi^{1/2}W_I^{1/2}}\frac{(-k\eta_{I})^{1/2-\alpha}}{k^{1/2}} ~, \\
h_{k*} &=& 
-2 ie^{i\pi\mu/2+i\pi/4} (1-\epsilon) 2^{\mu}\Gamma(\mu)\frac{H_*}{M_{\rm pl}}\frac{(-k\eta_*)^{3/2-\mu}}{k^{3/2}} ~, \label{eq:tens_mode_coeff}
\end{eqnarray}
and 
\begin{eqnarray}
u_{m}(x) = \frac{i\pi x^{m}}{2^{m}\Gamma(m)}H^{(1)}_{m}(x) ~, \label{eq:u_func}
\end{eqnarray}
with $\Gamma(m)$ and $H_m^{(1)}(x)$ being the Gamma function and Hankel function of the first kind, respectively. 
Here, the subscript $\ast$ denotes the value at a given time after the 
scale of interest exits the horizon,
$\eta = \eta_\ast$. 
With the asymptotic form of $u_m(x)$ at $-k\eta \to 0$, we find that on superhorizon scales ${\cal R}_k(\eta) \simeq {\cal R}_{k\ast}$ and $h_{k}(\eta) \simeq h_{k\ast}$. Properties of this function are described in appendix~\ref{appen:u}. Subscripts of this function are determined by the slow-roll parameters and the spectral index of running coupling as 
\begin{eqnarray}
\mu = \frac{3}{2} + \frac{\epsilon}{1-\epsilon} ~, \ \ 
\alpha = \frac{1}{2} + \frac{n}{1-\epsilon} ~, \ \ 
 \nu = \mu + \frac{\epsilon + \delta}{1-\epsilon} ~.
\end{eqnarray}

\subsection{Power spectrum of each mode} 

By use of the above equations, the power spectra of these perturbative quantities are immediately calculated: 
\begin{eqnarray}
\begin{split}
\Braket{\mathcal{R}({\bf k},\eta) \mathcal{R}({\bf k'},\eta')} 
&= (2\pi)^{3} \mathcal{R}_{k}(\eta)\mathcal{R}^{*}_{k}(\eta') \delta({\bf k} + {\bf k'}) ~ , \\
\Braket{A_i({\bf k},\eta) A_j({\bf k'},\eta')} 
&= (2\pi)^{3} v_{k}(\eta) v^{*}_{k}(\eta') P_{ij}(\hat{\bf k}) \delta({\bf k} + {\bf k'}) ~, \\
\Braket{h_{ij}({\bf k}, \eta) h_{kl}({\bf k'}, \eta') } 
&= (2\pi)^3 h_k(\eta) h_k^* (\eta') \Pi_{ij,kl} (\hat{\bf k})
\delta({\bf k} + {\bf k'}) ~, \label{eq:mode_power} 
\end{split}
\end{eqnarray}
with 
\begin{eqnarray}
\begin{split}
P_{ij}(\hat{\bf k}) &\equiv \sum_{\lambda = \pm 1} 
\epsilon_{i}^{(\lambda)} (\hat{\bf k}) \epsilon_{j}^{(- \lambda)} (\hat{\bf k}) 
 = \delta_{ij} - \hat{k}_i \hat{k}_j ~, \\
\Pi_{ij,kl}(\hat{\bf k}) &\equiv \sum_{\lambda = \pm 2} 
e_{ij}^{(\lambda)} (\hat{\bf k}) e_{kl}^{(- \lambda)} (\hat{\bf k}) ~.
\end{split}
\end{eqnarray}
One can find conventions of the polarization vector and tensor, $\epsilon_i^{(\lambda)}$ and $e_{ij}^{(\lambda)}$ in appendix~\ref{appen:polarization}.


The vector field can be decomposed into the electric and magnetic components as 
\begin{eqnarray}
\begin{split}
E_i({\bf k}, \eta) &= - A_i'({\bf k}, \eta) ~, \\ 
B_i({\bf k}, \eta) &= i \eta_{ijk} k_{j} A_{k}({\bf k},\eta) ~, 
\end{split}
\end{eqnarray}
where $~'~ \equiv \partial /\partial \eta $ denotes the derivative with respect to the conformal time and $\eta_{abc}$ is the 3D antisymmetric tensor normalized as $\eta_{123} = 1$.  Hence, the electric and magnetic power spectra outside the horizon at the end of inflation, namely, $-k \eta_I \ll 1$,
are respectively given by
\begin{eqnarray}
\begin{split}
\Braket{E_i({\bf k}, \eta_I) E^j({\bf k'}, \eta_I)} 
&= (2\pi)^3 \frac{P_E (k)}{2} P_{ij}(\hat{\bf k}) 
\delta\left( {\bf k} + {\bf k'} \right)  ~, \\ 
\Braket{B_i({\bf k}, \eta_I) B^j({\bf k'}, \eta_I)} 
&= (2\pi)^3 \frac{P_B(k)}{2} P_{ij}(\hat{\bf k}) 
\delta\left( {\bf k} + {\bf k'} \right)  ~, \\ 
P_E(k) &= 2 \frac{|v_k'(\eta_I)|^2}{a_I^4} ~, \\
P_B(k) &= 2 k^2 \frac{|v_k(\eta_I)|^2}{a_I^4} ~.
\end{split}
\end{eqnarray}

The magnetic component outside the horizon at the end of inflation re-enters the horizon at late stages and may become observed magnetic fields in case the vector fields can be considered as the direct seeds of the present magnetic fields. On the other hand, in such case 
the electric component quickly decays due to the large conductivity  (see, e.g., \cite{Martin:2007ue}). According to ref.~\cite{Motta:2012rn}, with the WMAP results \cite{Komatsu:2010fb}
and the assumption that $\epsilon = 0.01$ and the instantaneous reheating, 
the strength of the magnetic field at the present time can be calculated as 
\begin{eqnarray}
\frac{d}{d \ln k} \Braket{B_k^2} &=& \frac{k^3 P_B(k) a_I^4}{2\pi^2} \nonumber \\ 
&\simeq& 10^{-22.8 - 22.5 n_B} 
\frac{\Gamma\left(\frac{5-n_B}{2}\right)^2}{W_I \Gamma\left(\frac{5}{2}\right)^2} \left( \frac{k}{{\rm Mpc}^{-1}} \right)^{n_B} {\rm Gauss}^2 ~, \label{eq:mag_present}
\end{eqnarray}
where $n_B \equiv 4 - 2n/(1-\epsilon)$. This equation expresses that the magnetic field is drastically suppressed as $n_B$ increases. Hence, we can see that for $n = 2.1$, i.e., $n_B = -0.2$, and $W_I = 1$, the strongest magnetic field is produced as $B_{1 \rm Mpc} \simeq 10^{-9} {\rm Gauss}$. This value is comparable to upper bounds obtained from current CMB observations \cite{Finelli:2008xh, Shaw:2010ea, Paoletti:2010rx, Paoletti:2012bb, Trivedi:2011vt, Shiraishi:2010yk, Shiraishi:2012rm, Shiraishi:2011dh, Shiraishi:2011fi} and hence we can expect that the electromagnetic-scalar and electromagnetic-tensor correlations also have detectable impacts on CMB fluctuations. Thus, in section \ref{sec:cmb}, we focus on the analysis of CMB signals for $n_B = -0.2$, i.e., $n = 2.1$ and in order to treat more generic case we also study the effect of the electric component without considering the dilution of it.  

\section{Primordial cross-bispectra}\label{sec:bispectrum}

In this section, we calculate the primordial cross-bispectra between two electromagnetic fields and one curvature perturbation ($\Braket{{\cal R} EE}$ and $\Braket{{\cal R} BB}$), and those between two electromagnetic fields and one gravitational wave ($\Braket{h EE}$ and $\Braket{h BB}$)
\footnote{We do not deal with $\Braket{{\cal R} EB}$ and $\Braket{h EB}$ because these correlations do not generate the CMB signals as seen in section~\ref{sec:cmb}.}. 

As a powerful tool to compute the higher-order correlation in the inflationary Universe, we utilize the in-in formalism \cite{Maldacena:2002vr, Weinberg:2008hq}. In this formalism, the expectation value of the time-dependent operator in the interaction picture, $O(t)$, is given by 
\begin{eqnarray}
\Braket{O(t)} = \Braket{0| \left( T e^{- i \int H_{\rm int}(t') dt' } \right)^\dagger O(t) T e^{- i \int H_{\rm int}(t') dt' }  |0} ~,
\end{eqnarray}
where $T$ and $H_{\rm int}(t)$ are the time-ordering operator and interaction Hamiltonian, respectively. To compute the tree-level bispectra of primordial fluctuations, it is only necessary to use the first-order expression as
\begin{eqnarray}
\Braket{O(\eta)} = -2 {\rm Im} \int_{-\infty}^\eta d \eta' \Braket{0| : H_{\rm int}(\eta') : O(\eta) |0} ~, \label{eq:in-in_form}
\end{eqnarray}
where $:~:$ denotes the normal product. In the following discussion, we equate $\Braket{{\cal R}EE}$, $\Braket{{\cal R}BB}$, $\Braket{hEE}$ and $\Braket{hBB}$ to $\Braket{O(\eta)}$.

\subsection{Electromagnetic-scalar bispectra; $\Braket{{\cal R} EE}$ and $\Braket{{\cal R} BB}$}

Expanding the action (\ref{eq:EM_action}) to linear order in ${\cal R}$ leads to
\begin{eqnarray}
S_V^{({\cal R})} &=& \int d \eta d^3 x W(\varphi) 
\left\{
\left( 3 {\cal R} + \frac{{\cal R}'}{aH} \right) 
\left( \frac{1}{2} A_i'^2 - \frac{1}{4} F_{ij}^2 \right) \right. \nonumber \\ 
&& \left. \qquad\qquad\quad
- \left[ \left( {\cal R} + \frac{{\cal R}'}{aH} \right) A_i'^2 
- {\cal R} F_{ij}^2 - \frac{\partial_i {\cal R}}{aH} A_j' F_{ij} 
+ \epsilon \partial_i (\nabla^{-2} {\cal R}') A_j' F_{ij} 
  \right]
\right\} .
\end{eqnarray} 
Hence, the interaction Hamiltonian of the scalar part is derived as \cite{Motta:2012rn}
\begin{eqnarray}
\begin{split}
H_{\rm int}^{({\cal R})}(\eta_1) &=
\left[ \prod_{n=1}^3 \int \frac{d^3 {\bf k_n'}}{(2\pi)^3} \right] (2\pi)^3 \delta\left( \sum_{n=1}^3 {\bf k'_n} \right)
\times \left( - {W(\eta_1) \over 2} \right)  \\ 
&\qquad \times \Biggl\{
\left( {\cal R} ({\bf k_1'}, \eta_1) - {{\cal R}'({\bf k_1'}, \eta_1) \over a(\eta_1) H(\eta_1)} \right) \\
& \qquad\quad \times \left[
\delta^{ij} A_i'({\bf k_2}',\eta_1) A_j'({\bf k_3}',\eta_1)
 - (\delta^{ij}\delta^{k l} - \delta^{il} \delta^{jk})k_{2k}' k_{3 l}' A_i({\bf k_2}',\eta_1) A_j({\bf k_3}',\eta_1)\right]  \\
 & \qquad \quad 
 - {{\cal R}({\bf k_1}', \eta_1) \over a(\eta_1) H(\eta_1)}
 k_{1k}' \left[ k_{3l}'
 (\delta^{ij}\delta^{kl}- \delta^{il} \delta^{jk}) A_i'({\bf k_2}',\eta_1) A_j({\bf k_3}',\eta_1)
 + \left( 2 \leftrightarrow 3 \right)
 \right]\Biggr\}~.\label{eq:Hint_scal}
\end{split}
\end{eqnarray} 
Here, we have neglected the terms of order $\epsilon$. Note that this is equivalent to the interaction Hamiltonian in Ref.~\cite{Jain:2012vm} containing the total time derivative term. Substituting equation~(\ref{eq:Hint_scal}) into equation~(\ref{eq:in-in_form}) and overcoming somewhat complicated calculation, one obtain the explicit form of the electromagnetic-scalar bispectrum ($Y = E, B$) as
\begin{eqnarray}
\Braket{{\cal R}({\bf k_1},\eta) Y^i({\bf k_2}, \eta) Y_j({\bf k_3},\eta)}
&=& (2\pi)^{3}\delta\left(\sum_{n=1}^3 {\bf k_n} \right) \nonumber \\ 
&&\times 
 {\cal A}^{({\cal R})}(k_1, k_2, k_3, \eta) 
\left[ \sum_{n=1}^4 K_{n}^{{\cal R}YY} \mathcal{N}^{Y (n)}_{ij} \right]
 ~,
\label{eq:EM_scal}
\end{eqnarray}
where 
\begin{eqnarray}
{\cal A}^{({\cal R})}(k_1, k_2, k_3, \eta) &=& \frac{|{\cal R}_{k_{1}*}|^{2}|v_{k_{2}*}|^{2}|v_{k_{3}*}|^{2}}{a^{4}(\eta)}W_{I}(-k_1 \eta_I)^{\frac{2n}{1-\epsilon}} ~.
\end{eqnarray}
$K^{{\cal R}YY}$'s and ${\cal N}_{ij}^{Y}$'s involve the time integrals and angular dependence arising from contractions, which are respectively given by 
\begin{eqnarray}
\begin{split}
K_{1}^{{\cal R}EE}(k_1,k_2,k_3) &= - 4k_{1} ~ {\rm Im} ~
u_\nu^* (-k_1 \eta)  u_\alpha'^* (-k_2 \eta)  u_\alpha'^* (-k_3 \eta) \\ 
&\quad \times 
 \int^{\infty}_{\tau}d\tau_{1} \tau_1^{-\frac{2n}{1-\epsilon}} 
\left( 1+(1-\epsilon)\tau_{1}\frac{d}{d\tau_{1}} \right) 
u_{\nu}(\tau_{1}) 
\frac{d}{d\tau_{1}}u_{\alpha}(x_{2}\tau_{1}) 
\frac{d}{d\tau_{1}}u_{\alpha}(x_{3}\tau_{1})~, \\ 
K_{2}^{{\cal R}EE} (k_1,k_2,k_3)&= 2\frac{k_{2}k_{3}}{k_{1}} ~{\rm Im}~ 
u_\nu^* (-k_1 \eta)  u_\alpha'^* (-k_2 \eta)  u_\alpha'^* (-k_3 \eta) \\ 
&\quad\times 
\int^{\infty}_{\tau} d\tau_{1} \tau_1^{-\frac{2n}{1-\epsilon}} 
\left( 1+(1-\epsilon)\tau_{1}\frac{d}{d\tau_{1}}\right) 
u_{\nu}(\tau_{1})
u_{\alpha}(x_{2}\tau_{1}) 
u_{\alpha}(x_{3}\tau_{1}) ~, \\
K_{3}^{{\cal R}EE} (k_1,k_2,k_3)&= 2\frac{k_{2}k_{3}}{k_{1}} ~{\rm Im}~
u_\nu^* (-k_1 \eta)  u_\alpha'^* (-k_2 \eta)  u_\alpha'^* (-k_3 \eta) \\ 
&\quad\times 
\int^{\infty}_{\tau} 
d\tau_{1} \tau_1^{1-\frac{2n}{1-\epsilon}} (1-\epsilon)  u_{\nu}(\tau_{1}) 
u_{\alpha}(x_{2}\tau_{1}) \frac{d}{d\tau_{1}}u_{\alpha}(x_{3}\tau_{1}) ~, \\
K_{4}^{{\cal R}EE} (k_1,k_2,k_3)&= K_{3}^{{\cal R}EE}(k_1,k_3,k_2)~, \\ 
K_{1}^{{\cal R}BB} (k_1,k_2,k_3) &= -4k_{1}k_{2}k_{3} 
~ {\rm Im} ~
u_\nu^* (-k_1 \eta)  u_\alpha^* (-k_2 \eta)  u_\alpha^* (-k_3 \eta) \\ 
&\quad\times 
 \int^{\infty}_{-k_1 \eta}d\tau_{1} \tau_{1}^{- \frac{2n}{1-\epsilon}} 
\left( 1+(1-\epsilon)\tau_{1}\frac{d}{d\tau_{1}} \right)u_{\nu}(\tau_{1}) 
\frac{d}{d\tau_{1}}u_{\alpha}(x_{2}\tau_{1}) 
\frac{d}{d\tau_{1}}u_{\alpha}(x_{3}\tau_{1})~, \\ 
K_{2}^{{\cal R}BB}(k_1,k_2,k_3) &= 2\frac{k^{2}_{2}k^{2}_{3}}{k_{1}} 
~{\rm Im}~ 
u_\nu^* (-k_1 \eta)  u_\alpha^* (-k_2 \eta)  u_\alpha^* (-k_3 \eta) \\ 
&\quad\times 
\int^{\infty}_{-k_1 \eta} d\tau_{1} \tau^{- \frac{2n}{1-\epsilon}}_{1} 
\left( 1+(1-\epsilon)\tau_{1}\frac{d}{d\tau_{1}}\right) u_{\nu}(\tau_{1})
u_{\alpha}(x_{2}\tau_{1}) 
u_{\alpha}(x_{3}\tau_{1}) ~, \\
K_{3}^{{\cal R}BB}(k_1,k_2,k_3) &= 2\frac{k^{2}_{2}k^{2}_{3}}{k_{1}} ~{\rm Im}~
u_\nu^* (-k_1 \eta)  u_\alpha^* (-k_2 \eta)  u_\alpha^* (-k_3 \eta) \\ 
&\quad\times 
\int^{\infty}_{-k_1 \eta} 
d\tau_{1} \tau^{1-\frac{2n}{1-\epsilon}}_{1} (1-\epsilon) 
u_{\nu}(\tau_{1}) 
u_{\alpha}(x_{2}\tau_{1}) \frac{d}{d\tau_{1}}u_{\alpha}(x_{3}\tau_{1})  ~, \\
K_{4}^{{\cal R}BB}(k_1,k_2,k_3) &= K_{3}^{{\cal R}BB}(k_1,k_3,k_2)  ~, \label{eq:time_int_scal}
\end{split}
\end{eqnarray} 
and 
\begin{eqnarray}
\begin{split}
{\cal N}_{ij}^{E (1)}
&=  -\frac{1}{2} P_{ik}(\hat{\bf k_2}) P_{jk}(\hat{\bf k_3}) 
~, \\ 
{\cal N}_{ij}^{E (2)} 
&= - \hat{\bf k_2} \cdot \hat{\bf k_3} P_{ik}(\hat{\bf k_2}) P_{jk}(\hat{\bf k_3}) + 
 \hat{k_2}_l \hat{k_3}_k P_{ik}(\hat{\bf k_2}) P_{jl}(\hat{\bf k_3})
~, \\
{\cal N}_{ij}^{E (3)} 
&=  - \left( \frac {k_2}{k_3} + \hat{\bf k_2} \cdot \hat{\bf k_3} \right) 
P_{ik}(\hat{\bf k_2}) P_{jk}(\hat{\bf k_3}) 
+ \hat{k_{2}}_l \hat{k_{3}}_k 
P_{ik}(\hat{\bf k_2})  P_{jl}(\hat{\bf k_3})  ~, \\ 
{\cal N}_{ij}^{E (4)} 
&= - \left( \frac{k_3}{k_2} + \hat{\bf k_2} \cdot \hat{\bf k_3} \right) 
P_{ik}(\hat{\bf k_2})  P_{jk}(\hat{\bf k_3}) 
+ \hat{k_2}_l \hat{k_{3}}_k 
P_{ik}(\hat{\bf k_2}) P_{jl}(\hat{\bf k_3}) ~, \\
{\cal N}^{B (1)}_{ij} 
&= \frac{1}{2} 
\left( \hat{\bf k_2} \cdot \hat{\bf k_3} \delta_{ij} - \hat{k_2}_{j} \hat{k_3}_{i} \right) ~, 
\\
{\cal N}^{B (2)}_{ij} 
&= (\hat{\bf k_2} \cdot \hat{\bf k_3})^{2} \delta_{ij} 
- \hat{\bf k_2} \cdot \hat{\bf k_3} \hat{k_2}_{j} \hat{k_3}_{i} 
+ \eta_{imn}\hat{k_2}_{m}\hat{k_3}_{n} \eta_{jkl}\hat{k_2}_{k}\hat{k_3}_{l}~, \\ 
{\cal N}^{B (3)}_{ij} 
&= \left( \hat{\bf k_2} \cdot \hat{\bf k_3} +\frac{k_{2}}{k_{3}}\right) 
\left( \hat{\bf k_2} \cdot \hat{\bf k_3} \delta_{ij} 
- \hat{k_2}_{j} \hat{k_3}_{i} \right) 
+ \eta_{imn}\hat{k_3}_m \hat{k_2}_n \eta_{jkl} \hat{k_3}_{k}\hat{k_2}_{l} ~, \\
{\cal N}^{B (4)}_{ij} 
&= 
\left( \hat{\bf k_2} \cdot \hat{\bf k_3} +\frac{k_{3}}{k_{2}}\right) 
\left( \hat{\bf k_2} \cdot \hat{\bf k_3} \delta_{ij} - \hat{k_2}_{j} \hat{k_3}_{i} \right) 
+ \eta_{imn}\hat{k_2}_m \hat{k_3}_n \eta_{jkl} \hat{k_2}_{k}\hat{k_3}_{l}
~.
\end{split}
\end{eqnarray}
Here, $\tau_1 \equiv -k_1 \eta_{1}, x_2 \equiv k_2 / k_1$ and $x_3 \equiv k_3 / k_1$. Note that unlike ref.~\cite{Motta:2012rn}, in equation~(\ref{eq:EM_scal}), we do not perform the contraction between electromagnetic fields. It is because off-diagonal components of the electromagnetic-scalar bispectrum create significant signals on CMB anisotropies.

To see behaviors of these cross-bispectra at the end of inflation on superhorizon scales ($-k \eta_I \ll 1$), let us focus on two specific cases: $n = \pm 2$, i.e., $\alpha = 5/2$ and $-3/2$. 
Then, explicit forms of $K^{{\cal R}YY}$'s are given in appendix~\ref{appen:integral}. Taking into account the $\eta_I$-dependence of $K^{{\cal R}YY}$'s and $\cal A^{({\cal R})}$, we can find a fact: 
\begin{eqnarray}
\begin{split}
& \Braket{{\cal R}EE}_{n = 2} \propto (-k \eta_I)^{-2}~, \ \  
\Braket{{\cal R}BB}_{n = 2} \propto (-k \eta_I)^{-4} ~,  \\ 
& \Braket{{\cal R}EE}_{n = -2} \propto (-k \eta_I)^{-4} ~, \ \ 
\Braket{{\cal R}BB}_{n = -2} \propto (-k \eta_I)^{-2} ~.  
\end{split} 
\end{eqnarray}
This implies that the electric (magnetic) contribution dominates over the cross-bispectrum for $n = -2$ ($n = 2$). Such inverted behavior between $\Braket{{\cal R}EE}$ and $\Braket{{\cal R}BB}$ under the sign reversal of $n$ may be observed for any $n$. 
 
\subsection{Electromagnetic-tensor bispectra; $\Braket{hEE}$ and $\Braket{hBB}$}

In the same manner as the scalar case, expanding the action (\ref{eq:EM_action}) to linear order in $h_{ij}$ as 
\begin{eqnarray}
S_V^{(h)} =
-\int d\eta d^{3}x \frac{W(\varphi)}{2}\left[ h^{ij}A'_{i}A'_{j}-h^{ij}\delta^{kl}(\partial_{i}A_{k}\partial_{j}A_{l}+\partial_{k}A_{i}\partial_{l}A_{j})+2h^{ij}\delta^{kl}\partial_{i}A_{k}\partial_{l}A_{j}\right]~, \nonumber \\ 
\end{eqnarray}
and transforming into the Fourier components, we gain the tensor-part interaction Hamiltonian: 
\begin{eqnarray}
H_{\rm int}^{(h)} (\eta_1)
&=&
\left[ \prod_{n=1}^3 \int \frac{d^3{\bf k_n'}}{(2\pi)^3}\right] 
(2\pi)^3 \delta\left(\sum_{n=1}^3 {\bf k_n'} \right)
 \frac{W(\eta_1)}{2}  \nonumber \\ 
&& \times 
\left\{ h^{ij}({\bf k_1'},\eta_1) A_i'({\bf k_2}',\eta_1)
A_j'({\bf k_3'},\eta_1) \right. \nonumber \\ 
&&\quad \left.  
+ \left[ \delta^{ij}h^{kl}({\bf k_1'},\eta_1)+\delta^{kl}h^{ij}({\bf k_1'},\eta_1) 
- 2 \delta^{il}h^{kj}({\bf k_1'},\eta_1)
\right] \right. \nonumber \\ 
&&\qquad \left. \times {k_2'}_k {k_3'}_l A_{i}({\bf k_2'},\eta_{1})A_{j}({\bf k_3'}, \eta_{1})\right\} ~.\label{eq:Hint_tens}
\end{eqnarray} 
Substituting this Hamiltonian into equation~(\ref{eq:in-in_form}), 
we have 
\begin{eqnarray}
\Braket{h_{ij}({\bf k_1},\eta) Y^{k}({\bf k_2},\eta) Y_{l}({\bf k_3},\eta)} 
&=& (2\pi)^{3}\delta\left(\sum_{n=1}^3 {\bf k_n}\right) \nonumber \\ 
&&\times {\cal A}^{(h)}(k_1, k_2, k_3, \eta) 
\left[ \sum_{n=1}^2 K_n^{hYY} {\cal N}^{Y (n)}_{ijkl}  \right] ~, \label{eq:EM_tens}
\end{eqnarray}
where 
\begin{eqnarray}
{\cal A}^{(h)}(k_1, k_2, k_3, \eta) 
\equiv \frac{|h_{k_{1}*}|^{2}|v_{k_{2}*}|^{2}|v_{k_{3}*}|^{2}}{a^{4}(\eta)} W_I (-k_{1}\eta_I)^{\frac{2n}{1-\epsilon}} ~. 
\end{eqnarray}
In the tensor case, the time integrals and angular-dependent parts are summarized as
\begin{eqnarray}
\begin{split}
K_1^{hEE} 
&\equiv 2 k_1 ~{\rm Im}~ 
u_\mu^* (-k_1 \eta)  u_\alpha'^* (-k_2 \eta)  u_\alpha'^* (-k_3 \eta) \\ 
&\quad\times 
 \int^{\infty}_{-k_1 \eta} d\tau_1 ~ \tau_1^{-\frac{2n}{1-\epsilon}} u_{\mu}(\tau_1)\frac{\partial}{\partial \tau_1}u_{\alpha}(x_{2}\tau_1)\frac{\partial}{\partial \tau_1}u_{\alpha}(x_{3}\tau_1) ~, \\
K_2^{hEE} 
&\equiv 2 \frac{k_{2}k_{3}}{k_{1}} ~{\rm Im}~ 
u_\mu^* (-k_1 \eta)  u_\alpha'^* (-k_2 \eta)  u_\alpha'^* (-k_3 \eta)  \\ 
&\quad\times 
\int^{\infty}_{- k_1 \eta} d\tau_1 ~\tau_1^{- \frac{2n}{1-\epsilon}} u_{\mu}(\tau_1) u_{\alpha}(x_{2}\tau_1) u_{\alpha}(x_{3}\tau_1)~, \\ 
K_1^{hBB} 
&\equiv 2 k_1 k_2 k_3 ~{\rm Im}~ 
u_\mu^* (-k_1 \eta) u_\alpha^* (-k_2 \eta)  u_\alpha^* (-k_3 \eta) \\ 
&\quad\times 
 \int^{\infty}_{-k_1 \eta} d\tau_1 ~ \tau_1^{-\frac{2n}{1-\epsilon}} u_{\mu}(\tau_1)\frac{\partial}{\partial \tau_1}u_{\alpha}(x_{2}\tau_1)\frac{\partial}{\partial \tau_1}u_{\alpha}(x_{3}\tau_1) ~, \\
K_2^{hBB} 
&\equiv 2 \frac{(k_{2}k_{3})^2}{k_{1}} ~{\rm Im}~ 
u_\mu^* (-k_1 \eta)  u_\alpha^* (-k_2 \eta)  u_\alpha^* (-k_3 \eta) \\ 
&\quad\times 
\int^{\infty}_{- k_1 \eta} d\tau_1 ~\tau_1^{-\frac{2n}{1-\epsilon}} u_{\mu}(\tau_1) u_{\alpha}(x_{2}\tau_1) u_{\alpha}(x_{3}\tau_1)~, \label{eq:time_int_tens}
\end{split}
\end{eqnarray} 
and 
\begin{eqnarray}
\begin{split}
{\cal N}_{ijkl}^{E (1)} 
&= - \Pi_{mn,ij}(\hat{\bf k_1}) P_{km}(\hat{\bf k_2}) P_{l n}(\hat{\bf k_3}) 
~, \\
{\cal N}_{ijkl}^{E (2)} 
&= -  \hat{k_2}_m \hat{k_3}_n \Pi_{mn, ij}(\hat{\bf k_1}) P_{kr}(\hat{\bf k_2}) P_{l r}(\hat{\bf k_3})
- \hat{\bf k_2} \cdot \hat{\bf k_3} \Pi_{mn, ij}(\hat{\bf k_1}) P_{k m}(\hat{\bf k_2}) P_{l n}(\hat{\bf k_3}) \\ 
&\quad + \Pi_{mn,ij}(\hat{\bf k_1}) 
\left[ P_{kr}(\hat{\bf k_2}) P_{l n}(\hat{\bf k_3}) 
\hat{k_2}_m \hat{k_3}_r + 
 P_{k n}(\hat{\bf k_2}) P_{l r}(\hat{\bf k_3}) \hat{k_3}_m \hat{k_2}_r
\right]~, \\ 
{\cal N}_{ijkl}^{B (1)}
&=  
\Pi_{mn, ij}(\hat{\bf k_1}) \eta_{krm}\eta_{l q n}\hat{k_2}_{r}\hat{k_3}_{q} 
~, \\
{\cal N}_{ijkl}^{B (2)}
&= 
\Pi_{mn,ij}(\hat{\bf k_1}) \delta^{qr}\hat{k_2}_{s}\hat{k_3}_{t} \\
&\quad\times 
\left( \eta_{ksq}\eta_{l t r}\hat{k_2}_{m}\hat{k_3}_{n} 
+ \eta_{ksm}\eta_{l t n}\hat{k_2}_{q}\hat{k_3}_{r} -\eta_{ksq}\eta_{l t n} 
\hat{k_2}_{m}\hat{k_3}_{r}-\eta_{ksn}\eta_{l t q}\hat{k_2}_{r}\hat{k_3}_{m}\right) ~.
\end{split}
\end{eqnarray}

In the same manner as the scalar case, we analyze the cases for $n = \pm 2$. From the combination of analytic expressions of $K^{hYY}$'s described in appendix~\ref{appen:integral} and ${\cal A}^{(T)}$, we find the scaling relations as 
\begin{eqnarray}
\begin{split}
& \Braket{hEE}_{n = 2} \propto (k \eta_I)^{-2} ~, \ \ 
\Braket{hBB}_{n = 2} \propto (k \eta_I)^{-4} ~, \\  
& \Braket{hEE}_{n = -2} \propto (k \eta_I)^{-4} ~, \ \
 \Braket{hBB}_{n = - 2} \propto (k \eta_I)^{-2} ~. 
\end{split}
\end{eqnarray}
This dependence is identical to the scalar counterpart; hence for positive (negative) $n$, magnetic (electric) part dominates over the cross-bispectrum.  


\section{CMB power spectra}\label{sec:cmb} 

In this section, we discuss impacts of the electromagnetic-scalar (\ref{eq:EM_scal}) and electromagnetic-tensor (\ref{eq:EM_tens}) bispectra on CMB fluctuations. Then, notice that we do not deal with CMB bispectra but CMB power spectra because the CMB fluctuation arise quadratically from electromagnetic fields. 

\subsection{Formulation}

CMB anisotropies are quantified by one intensity ($X = {\cal I}$) and two linear polarization (${\cal E}, {\cal B}$) fields for the scalar ($Z = S$), vector ($V$) and tensor ($T$) modes \footnote{Here, we neglect the circular polarization.}. These are expanded by the spherical harmonics as
\begin{eqnarray}
\frac{\Delta X^{(Z)}(\hat{\bf n})}{X^{(Z)}} = \sum_{\ell m} 
a_{X, \ell m}^{(Z)} Y_{\ell m}(\hat{\bf n})~. 
\end{eqnarray}
Then, each coefficient is expressed as \cite{Shiraishi:2010sm, Shiraishi:2012bh}
\begin{eqnarray}
\begin{split}
a^{(Z)}_{X, \ell m} &= 4\pi (-i)^\ell 
\int \frac{k^2 dk}{(2\pi)^3} {\cal T}_{X, \ell}^{(Z)}(k) 
\sum_{\lambda} [{\rm sgn}(\lambda)]^{\lambda+x} \xi^{(\lambda)}_{\ell m}(k)~, \\
\xi^{(\lambda)}_{\ell m}(k) &= \int d^2 \hat{\bf k} {}_{-\lambda}Y_{\ell m}^*(\hat{\bf k})  
\xi^{(\lambda)} ({\bf k})~, \label{eq:alm_form}
\end{split}
\end{eqnarray}
where $\lambda$ denotes the helicity of each perturbation: $\lambda = 0$ ($Z = S$), $\pm 1$ ($V$) and $\pm 2$ ($T$), and $x$ discriminates the parity of each field: $x = 0$ ($X= {\cal I},{\cal E}$) and $1$ (${\cal B}$), respectively. $\xi^{(\lambda)}$ expresses the initial perturbation as $\xi^{(0)} \equiv {\cal R}$ and $\xi^{(\pm 2)} \equiv h^{(\pm 2)} = \frac{1}{2} O_{ij}^{(\mp 2)} h_{ij}$, and ${\cal T}_{X, \ell}^{(Z)}$ is the transfer function derived from the line-of-sight integral. 

As seen in equation~(\ref{eq:alm_form}), CMB anisotropies depend strongly on the magnitude of the initial fluctuations. Electromagnetic parts of the vector field also affect CMB anisotropies via the primordial perturbations as follows and those can generate both the scalar and the tensor modes. 
If electromagnetic fields exist in the radiation-dominated era, their anisotropic stresses act as sources of scalar and tensor metric perturbations. Due to this, on superhorizon scales logarithmically-growing metric perturbations arise prior to neutrino decoupling. After this, however, neutrino anisotropic stresses emerge and compensate for electromagnetic ones; therefore the enhancement of metric perturbations ceases. Consequently, we have \cite{Shaw:2009nf, Shiraishi:2012rm}
\begin{eqnarray}
\begin{split}
\xi_A^{(0)}({\bf k}) &\approx 
R_\gamma \ln\left(\frac{\eta_\nu}{\eta_I}\right) 
\frac{3}{2} O_{ij}^{(0)}(\hat{\bf k}) \Pi_{A ij}({\bf k})
 ~,\\
\xi_A^{(\pm 2)}({\bf k}) &\approx 
6 R_\gamma \ln\left(\frac{\eta_\nu}{\eta_I}\right)  
\frac{1}{2}O_{ij}^{(\mp 2)}(\hat{\bf k}) \Pi_{A ij}({\bf k})
~, \label{eq:ini_xi_pmf_ST}
\end{split}
\end{eqnarray}
where $\eta_\nu$ is the conformal time of neutrino decoupling, $R_\gamma \approx 0.6$ is the ratio of the energy density between photons and all relativistic particles, and a subscript $A$ denotes the quantity originated from electric and magnetic parts of the vector field. $\Pi_{A ij}$ means the time-independent energy momentum tensor of the residual electromagnetic field after the end of inflation as 
\begin{eqnarray}
\Pi^i_{A j}({\bf k}) 
&=& - \frac{W_I}{4\pi \rho_{\gamma}(\eta)} \int \frac{d^3 {\bf k'}}{(2 \pi)^3} 
\sum_{Y = E, B} Y^i({\bf k'}, \eta) Y_j({\bf k} - {\bf k'}, \eta) \nonumber \\
&=& - \frac{W_I a_I^4}{4\pi \rho_{\gamma,0}} \int \frac{d^3 {\bf k'}}{(2 \pi)^3} 
\sum_{Y = E, B} Y^i({\bf k'}, \eta_I) Y_j({\bf k} - {\bf k'}, \eta_I)~, \label{eq:ani_pmf}
\end{eqnarray}
with $\rho_{\gamma}$ and $\rho_{\gamma, 0}$ being the photon energy density and its present value, respectively. Note that due to $Y^i \propto W_I^{-1/2}$, $\Pi^i_{A j}$ is independent of $W_I$. This induces the absence of $W_I$ in $\xi_A^{(\lambda)}$. 
The metric perturbations outside the horizon behave as initial conditions of the CMB anisotropies of the scalar and tensor modes. Strictly speaking, electromagnetic fields also urge the modification of the transfer function. However, this change is negligible at large scales \cite{Shaw:2009nf} and hence we use the transfer functions without depending on electromagnetic fields \cite{Zaldarriaga:1996xe, Hu:1997hp, Shiraishi:2012bh} in our numerical calculation. 

On the other hand, electromagnetic fields also generate the CMB fluctuation of the vector mode because the vector-mode anisotropic stress equates to the Lorentz force and this supports the growth of the vorticity at recombination. Thus, the electromagnetic vector mode produces characteristic transfer function \cite{Mack:2001gc, Lewis:2004ef, Shiraishi:2011fi}. As the initial vector-mode perturbation, we adopt the fluctuation of the electromagnetic anisotropic stress as \cite{Shiraishi:2012rm, Shiraishi:2012sn, Shiraishi:2012bh}
\begin{eqnarray} 
\xi^{(\pm 1)}_A({\bf k}) \approx 
\frac{1}{2} O_{ij}^{(\mp 1)}(\hat{\bf k}) \Pi_{A ij}({\bf k}) ~. \label{eq:ini_xi_pmf_V}
\end{eqnarray}
Equations~(\ref{eq:ini_xi_pmf_ST}), (\ref{eq:ani_pmf}) and (\ref{eq:ini_xi_pmf_V}) imply the quadratic dependence of the initial perturbations on electromagnetic fields. This means that $\Braket{{\cal R}EE}, \Braket{{\cal R}BB}, \Braket{hEE}$ and $\Braket{hBB}$, which are given by equations~(\ref{eq:EM_scal}) and (\ref{eq:EM_tens}), equate to the power spectra of the initial perturbations as $\Braket{\xi^{(\lambda_1)} \xi_A^{(\lambda_2)}}$, and therefore become sources of CMB power spectra. 

From here, let us focus on a formulation of the CMB power spectra, which is expressed as
\begin{eqnarray}
\Braket{\prod_{n=1}^2 a_{X_n, \ell_n m_n}^{(Z_n)}} 
&=&  \left[\prod_{n=1}^2 4\pi (-i)^{\ell_n} 
\int \frac{k_n^2 dk_n}{(2\pi)^3} {\cal T}_{X_n, \ell_n}^{(Z_n)}(k_n) 
\sum_{\lambda_n} [{\rm sgn}(\lambda_n)]^{\lambda_n + x_n} \right]  \nonumber \\ 
&&\times \Braket{\xi_{\ell_1 m_1}^{(\lambda_1)}(k_1) \xi_{A,\ell_2 m_2}^{(\lambda_2)}(k_2)}~. \label{eq:cmb_pow_form}
\end{eqnarray}
At first, we should reduce the initial angular power spectra obtained from equations~(\ref{eq:EM_scal}), (\ref{eq:EM_tens}), (\ref{eq:ini_xi_pmf_ST}), (\ref{eq:ani_pmf}) and (\ref{eq:ini_xi_pmf_V}) as
\begin{eqnarray}
\Braket{\xi_{\ell_1 m_1}^{(\lambda_1)}(k_1) \xi_{A,\ell_2 m_2}^{(\lambda_2)}(k_2)}
 &=& C'_{-\lambda_2} \left( - \frac{W_I a_I^4}{4\pi \rho_{\gamma, 0}} \right) (-1)^{\ell_2 + \lambda_2} 
\int d^2 \hat{\bf k_1} {}_{-\lambda_1}Y_{\ell_1 m_1}^*(\hat{\bf k_1}) 
 {}_{\lambda_2}Y_{\ell_2 m_2}^*(\hat{\bf k_1}) 
 \nonumber \\
&&\times \int k_2'^2 d k_2' \int k_3'^2 d k_3' 
 \int d^2 \hat{\bf k_2'} \int d^2 \hat{\bf k_3'}
 F_{\lambda_1 \lambda_2}({\bf k_1}, {\bf k_2'}, {\bf k_3'})  \nonumber \\
&&\times 
\delta( {\bf k_1} + {\bf k_2'}+{\bf k_3'} )
\frac{\delta(k_1 - k_2)}{k_1^2}~, \label{eq:xi_ang_power_form}
\end{eqnarray}
with 
\begin{eqnarray}
\begin{split}
F_{0, \lambda_2}({\bf k_1}, {\bf k_2'}, {\bf k_3'}) 
&\equiv
{\cal A}^{({\cal R})}(k_1, k_2', k_3', \eta_I) \\ 
&\quad\times
\sum_{Y = E, B}
 \sum_{n=1}^4 K_{n}^{{\cal R}YY}(k_1, k_2', k_3') O_{ij}^{(\lambda_2)}(\hat{\bf k_1}) 
\mathcal{N}^{Y (n)}_{ij}(\hat{\bf k_2'}, \hat{\bf k_3'})  ~, \\
F_{\pm 2, \lambda_2}({\bf k_1}, {\bf k_2'}, {\bf k_3'}) 
&\equiv
{\cal A}^{(h)}(k_1, k_2', k_3', \eta_I)  \\
&\quad \times 
\sum_{Y = E, B}
 \sum_{n=1}^2 K_{n}^{hYY}(k_1, k_2', k_3') 
\frac{1}{2} e_{ij}^{(-\lambda_1)}(\hat{\bf k_1}) 
O_{kl}^{(\lambda_2)}(\hat{\bf k_1}) 
\mathcal{N}^{Y (n)}_{ijkl}(\hat{\bf k_1}, \hat{\bf k_2'}, \hat{\bf k_3'}) ~, \\ 
C'_{-\lambda_2} &\equiv
\begin{cases}
\frac{3}{2} R_\gamma \ln \left( \eta_\nu / \eta_I \right)& (\lambda_2 = 0) \\
\frac{1}{2} & (\lambda_2 = \pm 1) \\
3 R_\gamma \ln \left( \eta_\nu / \eta_I \right) & (\lambda_2 = \pm 2)
\end{cases} ~.
\end{split}
\end{eqnarray}
Here, we have decomposed the delta function into
\begin{eqnarray}
\delta({\bf k_1} + {\bf k_2}) = \frac{\delta(k_1 - k_2)}{k_1^2} \delta(\hat{\bf k_1} + \hat{\bf k_2})~.
\end{eqnarray}
For performing the angular integrals in the Fourier space, it is convenient to expand all angular dependent parts by the spin spherical harmonics. Then, the delta function is given by
\begin{eqnarray}
\delta( {\bf k_1} + {\bf k_2'} + {\bf k_3'} ) 
&=& 8 \int_0^\infty y^2 dy 
\sum_{\substack{L_1 L_2 L_3 \\ M_1 M_2 M_3}} 
 (-1)^{\frac{L_1 + L_2 + L_3}{2}}
j_{L_1}(k_1 y) j_{L_2}(k_2' y) j_{L_3}(k_3' y)  \nonumber \\
&&\times 
Y_{L_1 M_1}^*(\hat{\bf k_1}) Y_{L_2 M_2}^*(\hat{\bf k_2'})  Y_{L_3 M_3}^*(\hat{\bf k_3'}) 
I_{L_1 L_2 L_3}^{0~0~0}
\left(
  \begin{array}{ccc}
  L_1 & L_2 & L_3 \\
  M_1 & M_2 & M_3 
  \end{array}
 \right) ~,
\end{eqnarray}
where $j_{L}(x)$ is the spherical Bessel function and the $I$ symbol is defined by
\begin{eqnarray}
I^{s_1 s_2 s_3}_{l_1 l_2 l_3}
\equiv \sqrt{\frac{(2 l_1 + 1)(2 l_2 + 1)(2 l_3 + 1)}{4 \pi}}
\left(
  \begin{array}{ccc}
  l_1 & l_2 & l_3 \\
  s_1 & s_2 & s_3
  \end{array}
 \right)~. \label{eq:I_sym}
\end{eqnarray} 
$F_{\lambda_1 \lambda_2}$'s also involve the angular dependence. Taking the contractions by use of the conventions shown in appendix~\ref{appen:polarization} and refs.~\cite{Shiraishi:2010kd, Shiraishi:2012bh}, we have:
\begin{eqnarray}
\begin{split}
F_{0, \lambda_2}
 &= C_{\lambda_2} {\cal A}^{({\cal R})} 
\sum_{L' L''} 
\left[ \sum_{Y = E,B} \sum_{n=1}^3 {\cal K}_{n}^{{\cal R}YY}(k_1, k_2', k_3') {\cal V}_{L' L''}^{Y (n)} \right] 
\\
&\quad\times \sum_{M M' M''} 
{}_{-\lambda_2}Y_{2 M}^*(\hat{\bf k_1})
Y_{L' M'}^*(\hat{\bf k_2'}) Y_{L'' M''}^*(\hat{\bf k_3'})
 \left(
  \begin{array}{ccc}
  2 & L' & L'' \\
  M & M' & M''
  \end{array}
 \right) 
~, \\
F_{\pm 2, \lambda_2}
 &= C_{\lambda_2} {\cal A}^{(h)}
\sum_{L L' L''} 
\left[ \sum_{Y = E,B} \sum_{n=1}^2 K_n^{hYY}(k_1, k_2', k_3') {\cal V}_{L L' L''}^{Y (n)} \right]  
I_{2~ 2~ L}^{-\lambda_1 \lambda_2 -s} \\
&\quad \times
(-1)^{s + L + L' + L''} 
\sum_{M M' M''} 
{}_{-s}Y_{L M}^*(\hat{\bf k_1})
Y_{L' M'}^*(\hat{\bf k_2'}) Y_{L'' M''}^*(\hat{\bf k_3'})
 \left(
  \begin{array}{ccc}
  L & L' & L'' \\
  M & M' & M''
  \end{array}
 \right) 
~, \label{eq:contraction}
\end{split}
\end{eqnarray}
where $s \equiv \lambda_2 - \lambda_1$ and 
\begin{eqnarray}
C_{\lambda_2} &\equiv&
\begin{cases}
-2 & (\lambda_2 = 0) \\
2\sqrt{3} \lambda_2 & (\lambda_2 = \pm 1) \\
2\sqrt{3} & (\lambda_2 = \pm 2)
\end{cases} ~.
\end{eqnarray}
New functions for $\lambda_1 = 0$ and $\pm 2$ are respectively given by 
\begin{eqnarray}
{\cal K}_1^{{\cal R}YY} &=& - \frac{1}{2} K_1^{{\cal R}YY} - \frac{k_2}{k_3} K_3^{{\cal R}YY} - \frac{k_3}{k_2} K_4^{{\cal R}YY} ~, \\
{\cal K}_2^{{\cal R}YY} &=& - K_2^{{\cal R}YY} - K_3^{{\cal R}YY} - K_4^{{\cal R}YY} ~, \\
{\cal K}_3^{{\cal R}YY} &=& -{\cal K}_2^{{\cal R}YY} ~, \\
{\cal V}_{L' L''}^{E (1)} 
&=&  -4 I_{2 1 1}^{0 1 -1} 
\left( \frac{4\pi}{3} \right)^3 I_{L' 1 1 }^{0 1 -1} I_{L'' 1 1 }^{0 1 -1} 
\left( \delta_{L', 0} + \delta_{L', 2} \right) 
\left( \delta_{L'', 0} + \delta_{L'', 2} \right) 
 \left\{
  \begin{array}{ccc}
  2 & L' & L'' \\
  1 & 1 & 1  
  \end{array}
 \right\}
 ~, \\
{\cal V}_{L' L''}^{E (2)} 
&=& 4 I_{2 1 1}^{0 1 -1} 
\left( \frac{4\pi}{3} \right)^4 
\sum_{p_2, p_3 = 0, 2} I_{p_2 1 1 }^{0 1 -1} I_{p_3 1 1 }^{0 1 -1} 
 I_{1 p_2 L'}^{0 0 0} I_{1 p_3 L''}^{0 0 0} 
 \left\{
  \begin{array}{ccc}
  2 & p_2 & p_3 \\
  1 & 1 & 1
  \end{array}
 \right\}
 \left\{
  \begin{array}{ccc}
  2 & L' & L'' \\
  1 & p_3 & p_2
  \end{array}
 \right\} ~, \\
{\cal V}_{L' L''}^{E (3)} 
&=& 4 I_{2 1 1}^{0 1 -1} 
\left( \frac{4\pi}{3} \right)^4 
\sum_{p_2, p_3 = 0, 2} I_{p_2 1 1 }^{0 1 -1} I_{p_3 1 1 }^{0 1 -1} 
 I_{1 p_2 L'}^{0 0 0} I_{1 p_3 L''}^{0 0 0} 
 \left\{
  \begin{array}{ccc}
  2 & L' & L'' \\
  1 & p_2 & 1 \\
  1 & 1 & p_3 
  \end{array}
 \right\}
 ~, \\
{\cal V}_{L' L''}^{B (1)} 
&=&  I_{2 1 1}^{0 1 -1} 
\left( \frac{4\pi}{3} \right)^2 \delta_{L', 1} \delta_{L'', 1} ~, \\
{\cal V}_{L' L''}^{B (2)} 
&=& - I_{2 1 1}^{0 1 -1} 
\left( \frac{4\pi}{3} \right)^3
 I_{1 1 L'}^{0 0 0} I_{1 1 L''}^{0 0 0} 
 \left\{
  \begin{array}{ccc}
  2 & L' & L'' \\
  1 & 1 & 1
  \end{array}
 \right\} ~, \\
{\cal V}_{L' L''}^{B (3)} 
&=& 
- 6 I_{2 1 1}^{0 1 -1} 
\left( \frac{4\pi}{3} \right)^3
 I_{1 1 L'}^{0 0 0} I_{1 1 L''}^{0 0 0} 
 \left\{
  \begin{array}{ccc}
  2 & L' & L'' \\
  1 & 1 & 1 \\
  1 & 1 & 1 
  \end{array}
 \right\} ~,
\end{eqnarray}
and 
\begin{eqnarray}
{\cal V}_{L L' L''}^{E (1)} 
&=& - 8\sqrt{3}
\left( I_{2 1 1}^{0 1 -1} \right)^2 
\left( \frac{4\pi}{3} \right)^4  
(-1)^{L + L' + L''} \nonumber \\ 
&&\times 
(\delta_{L', 0} + \delta_{L', 2}) (\delta_{L'', 0} + \delta_{L'', 2}) I_{L' 1 1}^{0 1 -1} I_{L'' 1 1}^{0 1 -1} 
 \left\{
  \begin{array}{ccc}
  L & L' & L'' \\
  2 & 1 & 1 \\
  2 & 1 & 1 
  \end{array}
 \right\} ~, \\
{\cal V}_{L L' L''}^{E (2)} 
&=& 8 \sqrt{3} 
\left( I_{2 1 1}^{0 1 -1} \right)^2 
\left( \frac{4\pi}{3} \right)^5 
\sum_{p_2, p_3 = 0, 2} I_{p_2 1 1}^{0 1 -1} I_{p_3 1 1}^{0 1 -1} 
 \left\{
  \begin{array}{ccc}
  2 & p_2 & p_3 \\
  1 & 1 & 1 
  \end{array}
 \right\} 
I_{1 p_2 L'}^{0 0 0 } I_{1 p_3 L''}^{0 0 0 } 
  \left\{
  \begin{array}{ccc}
  L & L' & L'' \\
  2 & p_2 & p_3 \\
  1 & 1 & 1 
  \end{array}
 \right\} \nonumber \\
&& + 8 \sqrt{3} 
\left( I_{2 1 1}^{0 1 -1} \right)^2 
\left( \frac{4\pi}{3} \right)^5 
\sum_{p_2, p_3 = 0, 2} I_{p_2 1 1}^{0 1 -1} I_{p_3 1 1}^{0 1 -1} 
 \left\{
  \begin{array}{ccc}
  L & p_2 & p_3 \\
  2 & 1 & 1 \\
  2 & 1 & 1 
  \end{array}
 \right\} \nonumber \\ 
&&\quad \times 
I_{1 p_2 L'}^{0 0 0 }  I_{1 p_3 L''}^{0 0 0 } 
(-1)^{L' + L''} 
  \left\{
  \begin{array}{ccc}
  L & L' & L'' \\
  1 & p_3 & p_2 
  \end{array}
 \right\} \nonumber \\ 
&& - 8\sqrt{3} 
\left( I_{2 1 1}^{0 1 -1} \right)^2 
\left( \frac{4\pi}{3} \right)^5 
\sum_{p_2, p_3}
I_{1 p_2 L'}^{-1 1 0} I_{1 1 p_2}^{1 0 -1 }  
(\delta_{L', 1} + \delta_{L', 3})
I_{1 p_3 L''}^{-1 1 0} I_{1 1 p_3}^{1 0 -1 }  
(\delta_{L'', 1} + \delta_{L'', 3})
\nonumber  \\ 
&&\quad \times 
\left[1 + (-1)^{p_2 + p_3} \right] 
\left\{
  \begin{array}{ccc}
  2 & p_2 & p_3 \\
  1 & 1 & 1 
  \end{array}
\right\}
\left\{
 \begin{array}{ccc}
 L & L' & L'' \\
 2 & p_2 & p_3 \\
 2 & 1 & 1
  \end{array}
\right\} ~, \\
{\cal V}_{L L' L''}^{B (1)} 
&=& - 2\sqrt{3} \left( I_{2 1 1}^{0 1 -1} \right)^2 
\left( \frac{4\pi}{3} \right)^3 6 
 \left\{
  \begin{array}{ccc}
  L & 1 & 1 \\
  2 & 1 & 1 \\
  2 & 1 & 1
  \end{array}
 \right\}
\delta_{L', 1} \delta_{L'', 1}
 ~, \\
{\cal V}_{L L' L''}^{B (2)} 
&=& 
- 2\sqrt{3} \left( I_{2 1 1}^{0 1 -1} \right)^2 
\left( \frac{4\pi}{3} \right)^4
 I_{1 1 L'}^{0 0 0} I_{1 1 L''}^{0 0 0} 
 \left\{
  \begin{array}{ccc}
  L & L' & L'' \\
  2 & 1 & 1 \\
  2 & 1 & 1 
  \end{array}
 \right\} \nonumber  \\
&& + 2\sqrt{3} \left( I_{2 1 1}^{0 1 -1} \right)^2 
\left( \frac{4\pi}{3} \right)^4 6 (-1)^L
 I_{1 1 L'}^{0 0 0} I_{1 1 L''}^{0 0 0} 
 \left\{
  \begin{array}{ccc}
  L & 1 & 1 \\
  2 & 1 & 1 \\
  2 & 1 & 1 
  \end{array}
 \right\}
\left\{
  \begin{array}{ccc}
  L & L' & L'' \\
  1 & 1 & 1  
  \end{array}
 \right\} \nonumber  \\
&&+ 2\sqrt{3} \left( I_{2 1 1}^{0 1 -1} \right)^2 
\left( \frac{4\pi}{3} \right)^5 4 
 \left( \delta_{L',0} + \delta_{L',2} \right)
I_{1 1 1}^{1 -1 0} I_{1 1 L''}^{0 0 0} \nonumber \\ 
&&\quad\times 
\sum_{L_p = 1,2} I_{L_p 1 1}^{1 0 -1}  
I_{L_p 1 L'}^{1 -1 0}
 \left\{
  \begin{array}{ccc}
  L & L_p & 1 \\
  2 & 1 & 1 \\
  2 & 1 & 1 
  \end{array}
 \right\}
\left\{
  \begin{array}{ccc}
  L & L' & L'' \\
  1 & 1 & L_p  
  \end{array}
 \right\} \nonumber \\
&& + 2\sqrt{3} \left( I_{2 1 1}^{0 1 -1} \right)^2 
\left( \frac{4\pi}{3} \right)^5 4 
 \left( \delta_{L'',0} + \delta_{L'',2} \right)
I_{1 1 1}^{1 -1 0} I_{1 1 L'}^{0 0 0} (-1)^L \nonumber \\ 
&&\quad\times 
\sum_{L_p = 1,2} I_{L_p 1 1}^{1 0 -1}  
I_{L_p 1 L''}^{1 -1 0}
 \left\{
  \begin{array}{ccc}
  L & L_p & 1 \\
  2 & 1 & 1 \\
  2 & 1 & 1 
  \end{array}
 \right\}
\left\{
  \begin{array}{ccc}
  L & L' & L'' \\
  1 & L_p & 1  
  \end{array}
 \right\} ~. 
\end{eqnarray}
For calculating the tensor mode, we have used equation~(\ref{eq:P_eta_expand}).
By the Wigner symbols, the integrals of these spin spherical harmonics reduce to 
\begin{eqnarray}
\begin{split}
\int d^2 \hat{\bf k_1} Y_{\ell_1 m_1}^* {}_{\lambda_2}Y_{\ell_2 m_2}^* {}_{-\lambda_2}Y_{2 M}^* Y_{L_1 M_1}^* 
&= \sum_{L'''M'''} I_{\ell_1 L_1 L'''}^{0~ 0~ 0} I_{\ell_2~ 2~ L'''}^{- \lambda_2 \lambda_2 0} \\
&\quad \times 
(-1)^{M'''}
 \left(
  \begin{array}{ccc}
  \ell_1 & L_1 & L''' \\
  m_1 & M_1 & M'''
  \end{array}
 \right)
 \left(
  \begin{array}{ccc}
  \ell_2 & 2 & L''' \\
  m_2 & M & -M'''
  \end{array}
 \right)
 ~, \\
\int d^2 \hat{\bf k_1} {}_{-\lambda_1}Y_{\ell_1 m_1}^* {}_{\lambda_2}Y_{\ell_2 m_2}^* {}_{-s}Y_{L M}^* Y_{L_1 M_1}^*  
&= \sum_{L''' M'''} I_{\ell_1 L_1 L'''}^{\lambda_1 0 -\lambda_1} I_{\ell_2~ L~ L'''}^{- \lambda_2 s \lambda_1} \\
&\quad\times
(-1)^{M'''}
 \left(
  \begin{array}{ccc}
  \ell_1 & L_1 & L''' \\
  m_1 & M_1 & M'''
  \end{array}
 \right)
 \left(
  \begin{array}{ccc}
  \ell_2 & L & L''' \\
  m_2 & M & -M'''
  \end{array}
 \right) ~, \\
\int d^2 \hat{\bf k_2'} Y_{L_2 M_2}^* Y_{L' M'}^* 
&= (-1)^{M'} \delta_{L_2, L'} \delta_{M_2, -M'} ~, \\
\int d^2 \hat{\bf k_3'} Y_{L_3 M_3}^* Y_{L'' M''}^* 
&= (-1)^{M''} \delta_{L_3, L''} \delta_{M_3, -M''} ~.
\end{split}
\end{eqnarray}
Furthermore, the summations of Wigner symbols over azimuthal quantum numbers result in 
\begin{eqnarray}
\begin{split}
&  
\sum_{M' M'' } 
\sum_{M_2 M_3} 
 \left(
  \begin{array}{ccc}
  L & L' & L'' \\
  M & M' & M''
  \end{array}
 \right)
\left(
  \begin{array}{ccc}
  L_1 & L_2 & L_3 \\
  M_1 & M_2 & M_3 
  \end{array}
 \right) 
(-1)^{M' + M''}
\delta_{L_2, L'} \delta_{M_2, -M'}
\delta_{L_3, L''} \delta_{M_3, -M''}  \\
&\quad\qquad\qquad = \frac{1}{2L+1} 
 (-1)^{M} \delta_{L_1, L} \delta_{M_1, -M} 
(-1)^{L + L' + L''} \delta_{L_2, L'} \delta_{L_3, L''} ~, \\
& 
\sum_{M''' M M_1} 
(-1)^{M + M'''}
 \left(
  \begin{array}{ccc}
  \ell_1 & L_1 & L''' \\
  m_1 & M_1 & M'''
  \end{array}
 \right)
 \left(
  \begin{array}{ccc}
  \ell_2 & L & L''' \\
  m_2 & M & -M'''
  \end{array}
 \right)
\delta_{L_1, L} \delta_{M_1, -M} \\
&\quad\qquad\qquad= \frac{(-1)^{\ell_1 + L + L'''}}{2 \ell_1 + 1} \delta_{L_1, L} (-1)^{m_1} \delta_{\ell_1, \ell_2} \delta_{m_1, -m_2}  ~.
\end{split}
\end{eqnarray}
Here, the scalar-mode counterpart corresponds to the case where $L = 2$. By the selection rules of Wigner symbols \cite{Shiraishi:2010kd, Shiraishi:2012bh}, we can simplify the summations over $L'''$ for $\lambda_1 = 0$ and $\pm 2$, respectively, as
\begin{eqnarray}
\begin{split}
\sum_{L''' }
I_{\ell_1 2 L'''}^{0~ 0~ 0} I_{\ell_1 2 L'''}^{- \lambda_2 \lambda_2 0} 
&=
\frac{5 (2 \ell_1 + 1)}{4 \pi} \delta_{\lambda_2, 0} ~, \\
 I_{2~2~L_1}^{-\lambda_1 \lambda_2 -s} 
\sum_{L''' } (-1)^{L'''} I_{\ell_1 L_1 L'''}^{\lambda_1 0 -\lambda_1} I_{\ell_1 L_1 L'''}^{- \lambda_2 s \lambda_1}
&= \delta_{\lambda_1, \lambda_2}
I_{2~2~L_1}^{-\lambda_1 \lambda_1 0} 
 (-1)^{\ell_1 + L_1} \sum_{L''' } \left( I_{\ell_1 L_1 L'''}^{\lambda_1 0 -\lambda_1} \right)^2 ~.
\end{split}
\end{eqnarray}
These results ensure that the couplings between different circular modes, such as $\Braket{\xi^{(0)} \xi_A^{(\pm 1, \pm 2)}}$, vanish. Due to this fact, the CMB power spectrum of the vector mode does not arise from the electromagnetic-scalar and electromagnetic-tensor bispectra. From the above treatments, the initial angular power spectra for $\lambda_1 = 0$ and $\pm 2$ (\ref{eq:xi_ang_power_form}) can respectively reduce to
\begin{eqnarray}
\Braket{\xi_{\ell_1 m_1}^{(0)}(k_1) \xi_{A,\ell_2 m_2}^{(\lambda_2)}(k_2)}
&=& (-1)^{m_1} \delta_{\ell_1, \ell_2} \delta_{m_1, -m_2} \delta_{\lambda_2, 0}
(-3) R_\gamma \ln \left( \eta_\nu \over \eta_I \right) 
\left(- \frac{a_I^4}{4\pi \rho_{\gamma, 0}} \right) 
(-1)^{\ell_1}  
 \nonumber \\
&&\times  
 \int_0^\infty y^2 dy
j_2(k_1 y) 
\frac{2}{\pi}  
\frac{\delta(k_1 - k_2)}{k_1^2} 
\sum_{L_2 L_3} (-1)^{\frac{2 + L_2 + L_3}{2}} I_{2 L_2 L_3}^{0~0~0}
  \nonumber \\
&&\times
\left[ \prod_{n=2}^3 \int k_n'^2 d k_n' j_{L_n}(k_n' y)  \right]
W_I {\cal A}^{({\cal R})}(k_1, k_2', k_3', \eta_I) \nonumber \\
&&\times \left[ \sum_{Y = E, B} \sum_{n=1}^3 {\cal K}_n^{{\cal R}YY}(k_1, k_2', k_3') {\cal V}_{L_2 L_3}^{Y (n)} \right] 
~, \label{eq:xi_S_ang_power} \\
\Braket{\xi_{\ell_1 m_1}^{(\lambda_1)}(k_1) \xi_{A,\ell_2 m_2}^{(\lambda_2)}(k_2)}
&=& (-1)^{m_1} \delta_{\ell_1, \ell_2} \delta_{m_1, -m_2} \delta_{\lambda_1, \lambda_2}
6\sqrt{3} R_\gamma \ln \left( \eta_\nu \over \eta_I \right) 
\left(- \frac{a_I^4}{4\pi \rho_{\gamma, 0}} \right) 
(-1)^{\ell_1}  
 \nonumber \\
&&\times
 \sum_{L_1 = 0,2,4}
I_{2~2~L_1}^{-\lambda_1 \lambda_1 0} 
\left[ \sum_{L''' } 
\left( I_{\ell_1 L_1 L'''}^{\lambda_1 0 -\lambda_1} \right)^2  \right]
\frac{1}{(2 L_1 + 1) (2 \ell_1 + 1)} 
 \nonumber \\
&&\times  
8 \int_0^\infty y^2 dy
j_{L_1}(k_1 y) 
\frac{\delta(k_1 - k_2)}{k_1^2} 
\sum_{L_2 L_3} (-1)^{\frac{L_1 + L_2 + L_3}{2}} I_{L_1 L_2 L_3}^{0~0~0}
  \nonumber \\
&&\times
\left[ \prod_{n=2}^3 \int k_n'^2 d k_n' j_{L_n}(k_n' y)  \right]
W_I {\cal A}^{(h)}(k_1, k_2', k_3', \eta_I) \nonumber \\
&&\times \left[ \sum_{Y = E, B} \sum_{n=1}^2 K^{hYY}_n(k_1, k_2', k_3') {\cal V}_{L_1 L_2 L_3}^{Y (n)} \right] 
~. \label{eq:xi_T_ang_power}
\end{eqnarray}
Note that $(-1)^{m_1} \delta_{\ell_1, \ell_2} \delta_{m_1, -m_2}$ enforces the rotational invariance of the initial and CMB power spectrum.

In order to convert into CMB power spectra, we substitute these equations into equation~(\ref{eq:cmb_pow_form}). As a result, we obtain final formulae for CMB power spectra of the scalar and tensor modes, respectively, as
\begin{eqnarray}
\Braket{\prod_{n=1}^2 a^{(Z_n)}_{X_n, \ell_n m_n}} &=& C_{X_1 X_2, \ell_1}^{(Z_1 Z_2)} (-1)^{m_1} \delta_{\ell_1, \ell_2} \delta_{m_1, -m_2} ~, \\
C_{X_1 X_2, \ell_1}^{(S Z_A)} 
&=& \delta_{S,Z_A}
 \delta_{x_2,0}
(-3) R_\gamma \ln \left( \eta_\nu \over \eta_I \right) 
\left(- \frac{a_I^4}{4\pi \rho_{\gamma, 0}} \right) 
 \nonumber \\
&&\times  
 \int_0^\infty y^2 dy 
\frac{2}{\pi} \int k_1^2 dk_1 j_2(k_1 y) 
{\cal T}_{X_1, \ell_1}^{(S)}(k_1) {\cal T}_{X_2, \ell_1}^{(S)}(k_1) 
  \nonumber \\
&&\times 
\sum_{L_2 L_3} (-1)^{\frac{2 + L_2 + L_3}{2}} I_{2 L_2 L_3}^{0~0~0}
\left[ \prod_{n=2}^3  \int \frac{k_n'^2 d k_n'}{2 \pi^2}  j_{L_n}(k_n' y) \right] \nonumber \\
&&\times W_I {\cal A}^{({\cal R})}(k_1, k_2', k_3', \eta_I) 
\left[ \sum_{Y = E, B} \sum_{n=1}^3 {\cal K}_{n}^{{\cal R}YY}(k_1, k_2', k_3') {\cal V}_{L_2 L_3}^{Y (n)} \right] 
 ~, \label{eq:cl_scal} \\
C_{X_1 X_2, \ell_1}^{(T Z_A)} 
&=& 
\delta_{T, Z_A} (\delta_{x_1,0} \delta_{x_2,0} + \delta_{x_1,1} \delta_{x_2,1})
6\sqrt{3} R_\gamma \ln \left( \eta_\nu \over \eta_I \right) 
\left(- \frac{a_I^4}{4\pi \rho_{\gamma, 0}} \right)  \nonumber \\
&&\times
 \sum_{L_1 = 0,2,4}
2 I_{2 2 L_1}^{-2 2 0} 
\left[ \sum_{L''' } 
\left( I_{\ell_1 L_1 L'''}^{2~ 0~ -2} \right)^2 \right] 
\frac{1}{(2 L_1 + 1) (2 \ell_1 + 1)} 
 \nonumber \\
&&\times  
8 \int_0^\infty y^2 dy 
\int k_1^2 dk_1 j_{L_1}(k_1 y) 
{\cal T}_{X_1, \ell_1}^{(T)}(k_1) {\cal T}_{X_2, \ell_1}^{(T)}(k_1) \nonumber \\
&&\times \sum_{L_2 L_3} (-1)^{\frac{L_1 + L_2 + L_3}{2}} I_{L_1 L_2 L_3}^{0~0~0}
\left[ \prod_{n=2}^3  \int \frac{k_n'^2 d k_n'}{2 \pi^2}  j_{L_n}(k_n' y) \right] 
\nonumber \\
&&\times  
W_I {\cal A}^{(h)}(k_1, k_2', k_3', \eta_I) 
\left[\sum_{Y = E, B} \sum_{n=1}^2 K^{hYY}_{n}(k_1, k_2', k_3') {\cal V}_{L_1 L_2 L_3}^{Y (n)} \right]~. \nonumber \\ 
\label{eq:cl_tens}
\end{eqnarray}
In the derivation of equation~(\ref{eq:cl_tens}), we have performed the summation over $\lambda_1 = \pm 2$ and $\lambda_2$ as 
\begin{eqnarray}
&& \left( \prod_{n=1}^2 \sum_{\lambda_n} [{\rm sgn}(\lambda_n)]^{\lambda_n + x_n} 
\right)
 \delta_{\lambda_1, \lambda_2} 
I_{2~2~L_1}^{-\lambda_1 \lambda_1 0} \left( I_{\ell_1 L_1 L'''}^{\lambda_1 0 -\lambda_1} \right)^2  \nonumber \\
&&\qquad\qquad= 
\begin{cases}
 2 I_{2 2 L_1}^{-2 2 0} 
\left( I_{\ell_1 L_1 L'''}^{2~ 0~ -2} \right)^2 & (x_1 + x_2 + L_1 = {\rm even}) \\
 0 & (x_1 + x_2 + L_1 = {\rm odd})
\end{cases} 
~.
\end{eqnarray}
From these expressions, we can see that due to ${\cal A}^{({\cal R},h)} \propto W_I^{-1}$, CMB power spectra are independent of $W_I$. Therefore, the behaviors and amplitudes of CMB power spectra are determined by only the spectral tilt of the running coupling, $n$, except some inflationary parameters.

\subsection{Numerical results}%

Here, we analyze CMB signals through the numerical computation of CMB power spectra, which are given by (\ref{eq:cl_scal}) and (\ref{eq:cl_tens}). Let us consider the case where the strength of the magnetic part is maximized without spoiling inflation, namely, $n = 2.1$. As mentioned in section~\ref{sec:inflation}, the CMB signals for this case may be enhanced to the level we can observe. 
Considering the standard single field slow-roll inflation, 
the slow-roll parameters are small, i.e., $\epsilon, \delta \ll 1$,
and  the tensor-to-scalar ratio denoted by $r$ can be related with the slow-roll parameter
as $r = 16 \epsilon$.
Neglecting the slow-roll corrections, we fix
 the parameters as $\alpha = 2.6$ and $\nu = \mu = 3/2$ and hence
 we have
\begin{eqnarray}
W_I {\cal A}^{({\cal R})}(k_1, k_2', k_3', \eta_I) 
&=& 224.575 \frac{H^{2}_*}{\epsilon M^{2}_{\rm pl}}H_I^{4.2}a_I^{0.2} k_1^{1.2} k_2'^{-5.2} k_3'^{-5.2} ~, \label{eq:calAS_n2.1} \\
W_I {\cal A}^{(h)}(k_1, k_2', k_3', \eta_I) &=& 
{r \over 4} W_I {\cal A}^{({\cal R})}(k_1, k_2', k_3', \eta_I)  ~. \label{eq:calAT_n2.1}
\end{eqnarray}

 For explicitly obtaining the CMB power spectra, we also need concrete forms of ${\cal K}^{{\cal R}YY}$'s and $K^{h YY}$'s. However, calculation of these for $\alpha = 2.6$ is so complicated that we here substitute the expressions for $\alpha = 5/2$. In appendix~\ref{appen:integral}, these analytical forms are provided and we found that ${\cal K}^{{\cal R}EE}$ and ${\cal K}^{hEE}$ are negligible compared with ${\cal K}^{{\cal R}BB}$ and ${\cal K}^{hBB}$ due to the suppression by $(- k \eta_I)^2$. In addition, we notice that in ${\cal K}^{{\cal R}BB}$'s and ${\cal K}^{hBB}$'s, only ${\cal K}_2^{{\cal R}BB}, {\cal K}_3^{{\cal R}BB}$ and $K_2^{hBB}$ involve the terms which grow logarithmically outside the horizon as $\ln(-\omega \eta_I)$ with $\omega \equiv k_1+k_2'+k_3'$. Considering the contributions on the interesting scales to calculate the CMB signals, namely, $10^{-6} \lesssim k \times {\rm Mpc} \lesssim 0.1$,
$\ln (- \omega \eta_I)$ is just identical to the e-folding number and it is about $- 60 \sim -50$. 
On the other hand, the other terms in ${\cal K}_2^{{\cal R}BB}, {\cal K}_3^{{\cal R}BB}$ and $K_2^{hBB}$ have the power-law dependence on $k$ and their coefficients are order of unity. Therefore, we believe that the logarithmic terms dominate over ${\cal K}_2^{{\cal R}BB}, {\cal K}_3^{{\cal R}BB}$ and $K_2^{hBB}$. Following this concept, we use such approximate forms as 
\begin{eqnarray}
\begin{split}
 {\cal K}_2^{{\cal R}BB} &= - {\cal K}_3^{{\cal R}BB} 
\simeq -\frac{160 k_2'^2 k_3'^2}{k_1} 
~, \\
K_2^{hBB} &\simeq 
\frac{40 k_2'^2 k_3'^2}{k_1} ~, \label{eq:calK_app} 
\end{split}
\end{eqnarray}
and neglect other ${\cal K}$'s and $K$'s.

 \begin{figure}[t]
  \begin{tabular}{cc}
    \begin{minipage}{0.5\hsize}
  \begin{center}
    \includegraphics[width=7.5cm,clip]{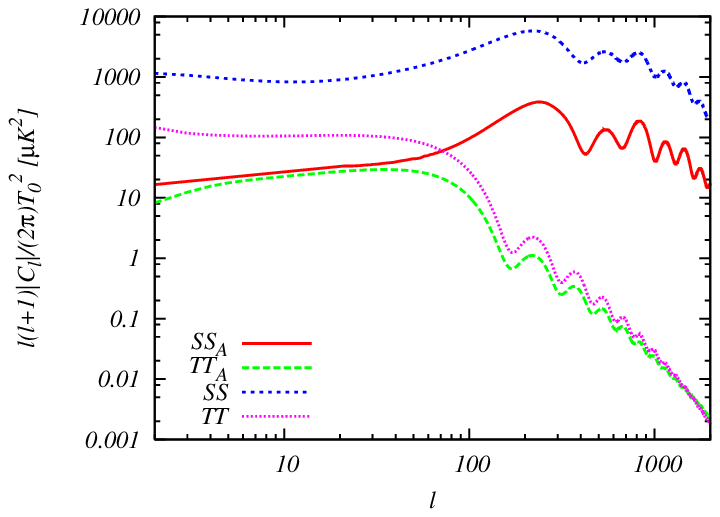}
  \end{center}
\end{minipage}
\begin{minipage}{0.5\hsize}
  \begin{center}
    \includegraphics[width=7.5cm,clip]{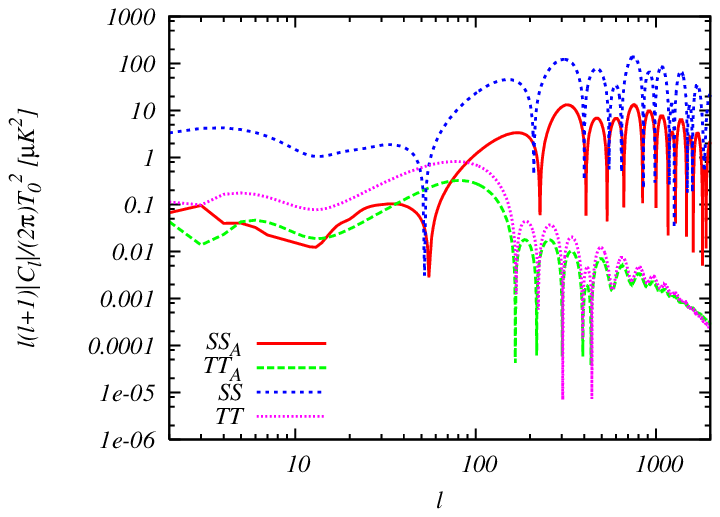}
  \end{center}
\end{minipage}
\\
    \begin{minipage}{0.5\hsize}
  \begin{center}
    \includegraphics[width=7.5cm,clip]{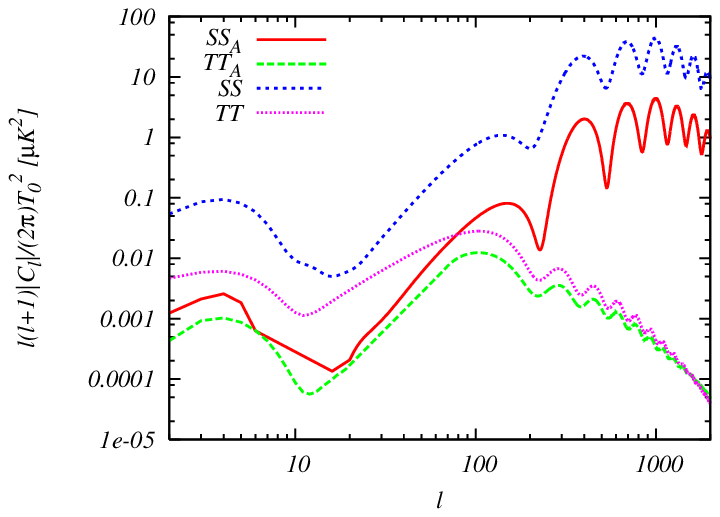}
  \end{center}
\end{minipage}
\begin{minipage}{0.5\hsize}
  \begin{center}
    \includegraphics[width=7.5cm,clip]{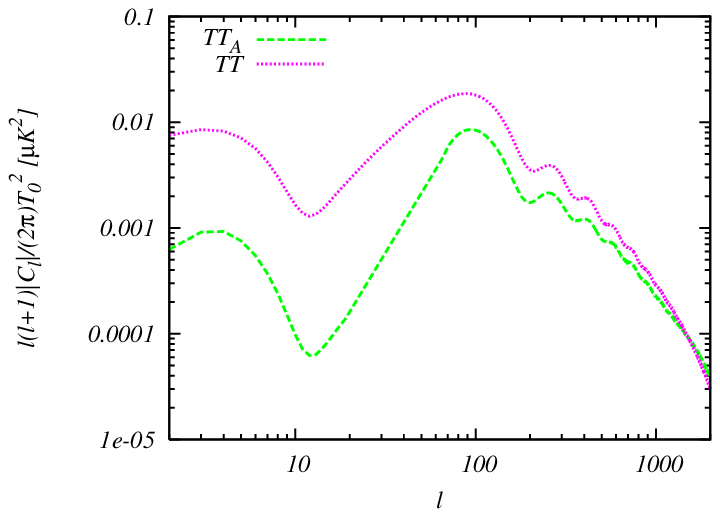}
  \end{center}
\end{minipage}
\end{tabular}
  \caption{Absolute values of CMB power spectra of the ${\cal II}$ (left top panel), ${\cal IE}$ (right top one), ${\cal EE}$ (left bottom one) and ${\cal BB}$ (right bottom one) modes. Red solid ($SS_A$), green dashed ($TT_A$), blue dotted ($SS$) and magenta dot-dashed ($TT$) curves correspond to the spectra sourced from the electromagnetic-scalar bispectra (\ref{eq:cl_scal}), electromagnetic-tensor correlation (\ref{eq:cl_tens}), scalar auto-correlation (\ref{eq:cl_scal_auto}) and tensor auto-correlation (\ref{eq:cl_tens_auto}), respectively. 
Here, we have adopted $H_I = 5 \times 10^{13} {\rm GeV}, a_I = 10^{-29}$ and $r = 16 \epsilon = 0.26$. 
The other cosmological parameters have been fixed to the mean values reported in ref.~\cite{Komatsu:2010fb}.} \label{fig:cl}
\end{figure}

Figure~\ref{fig:cl} shows the CMB ${\cal II}$ (left top panel), ${\cal IE}$ (right top one), ${\cal EE}$ (left bottom one) and ${\cal BB}$ (right bottom one) power spectra generated from $\Braket{{\cal R}YY}$ (red solid curves) and $\Braket{hYY}$ (green dashed ones) respectively given by equations~(\ref{eq:cl_scal}) and (\ref{eq:cl_tens}) (hereafter referred to as the $S S_A$ and $T T_A$ cases, respectively). For comparison, we also plot the CMB power spectra induced by the scalar (blue dotted curves) and tensor (magenta dot-dashed ones) auto-correlations, which are formulated as 
\begin{eqnarray}
C_{X_1 X_2, \ell_1}^{(S S)} &=& \frac{2}{\pi} \int k_1^2 dk_1 P_{\cal R}(k_1) {\cal T}_{X_1, \ell_1}^{(S)}(k_1) {\cal T}_{X_2, \ell_1}^{(S)}(k_1)  ~, \label{eq:cl_scal_auto} \\
C_{X_1 X_2, \ell_1}^{(T T )} &=& \frac{2}{\pi} \int k_1^2 dk_1 P_{h}(k_1) {\cal T}_{X_1, \ell_1}^{(T)}(k_1) {\cal T}_{X_2, \ell_1}^{(T)}(k_1)  
~, \label{eq:cl_tens_auto}
\end{eqnarray}
(hereafter referred to as the $SS$ and $TT$ cases, respectively). 
Here, $P_{\cal R}(k) = |{\cal R}_{k *}|^2$ and $P_h(k) = 2 |h_{k*}|^2$ are power spectra of primordial curvature perturbations and primordial gravitational waves on the superhorizon scales, respectively. To obtain these spectra, we used the modified version of the Boltzmann Code for Anisotropies in the Microwave Background (CAMB) \cite{Lewis:2004ef, Lewis:1999bs} and the Common Mathematical Library SLATEC \cite{slatec}. 

In this figure, comparing the red solid curves with the blue dotted ones, we can see that the overall behaviors of $C_\ell$'s for the $S S_A$ case are in good agreement with those for the $SS$ case. This may be due to the same transfer functions of both cases. As seen in equations~(\ref{eq:cl_scal}) and (\ref{eq:cl_scal_auto}), both cases have completely different dependence on multipoles. Nevertheless, the transfer functions strongly impact on the shapes of the CMB power spectra. Although the tensor modes are not as clear as the scalar modes, similar behaviors can be observed in the $T T_A$ and $TT$ cases. 

In contrast, there are differences of the amplitudes between the $S S_A$ and $S S$ cases. From this figure, we can see $C_{X_1 X_2, \ell}^{(S S_A)} / C_{X_1 X_2, \ell}^{(S S)} \sim 0.01$. This value can be understood by the naive estimation of the magnitudes of primordial perturbations. 
By using equations~(\ref{eq:mode_power}), (\ref{eq:mag_present}), (\ref{eq:calAS_n2.1}), and observational values of the cosmological parameters \cite{Komatsu:2010fb}, we gain $\Braket{\xi^{(0)} \xi^{(0)}} \sim 2.4 \times 10^{-9}$ and $\Braket{\xi^{(0)} \xi_A^{(0)}} \sim 3.3 \times 10^{-11}$. 
As seen in the previous sections, since $C_{X_1 X_2, \ell}^{(S S_A)}$ is independent of $W_I$, it can be believed that $C_{X_1 X_2, \ell}^{(S S_A)} / C_{X_1 X_2, \ell}^{(S S)} \sim \Braket{\xi^{(0)} \xi_A^{(0)}} / \Braket{\xi^{(0)} \xi^{(0)}}$ is a good approximation. Hence, we can get the consistent result. 
In the same manner, from equation~(\ref{eq:calAT_n2.1}), the magnitudes of primordial tensor perturbations are evaluated as 
$\Braket{\xi^{(\pm 2)} \xi^{(\pm 2)}} \sim 1.6 \times 10^{-10}$ and $\Braket{\xi^{(\pm 2)} \xi_A^{(\pm 2)}} \sim 1.3 \times 10^{-11}$. 
Therefore, we see that at large scales, in particular, a naive estimation, i.e., $C_{X_1 X_2, \ell}^{(T T_A)} / C_{X_1 X_2, \ell}^{(T T)} \sim \Braket{\xi^{(\pm 2)} \xi_A^{(\pm 2)}} / \Braket{\xi^{(\pm 2)} \xi^{(\pm 2)}} \sim 0.1$, is consistent with figure~\ref{fig:cl}. A fact that $C_{X_1 X_2, \ell}^{(T T_A)} / C_{X_1 X_2, \ell}^{(T T)}$ is about ten times larger than $C_{X_1 X_2, \ell}^{(S S_A)} / C_{X_1 X_2, \ell}^{(S S)}$ is due to the difference of the metric perturbations induced by electromagnetic fields between the scalar and tensor modes as shown in equation~(\ref{eq:ini_xi_pmf_ST}). Unfortunately, the $SS$ case dominate over the CMB power spectra and the $S S_A$ signals are subdominant components. Regardless of it, the above fact may help us to differentiate the $S S_A$ signals from the $S S$ ones, respectively. On the other hand, at small scales, the $T T_A$ spectra is comparable to the $TT$ spectra. This enhancement can be observed in the ${\cal BB}$ mode since there are no noise of the $SS$ spectra unlike in the ${\cal II}, {\cal IE}$ and ${\cal EE}$ modes.

Furthermore, interestingly, the ${\cal II}$ and ${\cal EE}$ modes for the $SS_A$ cases have negative signals. More precisely, the $SS_A$ signals are opposite in sign to the $SS$ ones in the ${\cal II}$, ${\cal IE}$ and ${\cal EE}$ modes. This is a consequence of the fact that the $SS_A$ signals are generated from the cross-bispectrum of two electromagnetic fields and one curvature perturbation and hence can take negative values. This may also become a clue as to the existence of the interaction between the scalar and vector fields (\ref{eq:EM_action}). 

There may remain a concern about the contributions of CMB power spectra sourced from $\Braket{YYYY}$. However, in the similar manner as the above discussion, from equation~(\ref{eq:ini_xi_pmf_ST}), these amplitudes are evaluated as 
\begin{eqnarray}
C_{X_1 X_2, \ell}^{(S_A S_A)} / C_{X_1 X_2, \ell}^{(S S)} \sim \Braket{\xi_A^{(0)} \xi_A^{(0)}} / \Braket{\xi^{(0)} \xi^{(0)}} \sim 10^{-4}~,
\end{eqnarray}
 and 
\begin{eqnarray}
C_{X_1 X_2, \ell}^{(T_A T_A)} / C_{X_1 X_2, \ell}^{(T T)} \sim \Braket{\xi_A^{(\pm 2)} \xi_A^{(\pm 2)}} / \Braket{\xi^{(\pm 2)} \xi^{(\pm 2)}} \sim 10^{-2}~,
\end{eqnarray}
 and thus we believe that total CMB signals do not drastically change.

\section{Summary and discussion}

In this paper, we studied the impacts of primordial cross-bispectra between the vector fields and metric perturbations on the CMB anisotropies. 
In the previous study \cite{Motta:2012rn}, only the cross-bispectrum between one curvature perturbation and two magnetic fields is analyzed. However, the kinetic term of the vector field (\ref{eq:EM_action}) also induces the cross-bispectra composed of electric fields and tensor perturbation. For the sake of completeness, we first formulated the primordial cross-bispectra: $\Braket{{\cal R}EE}, \Braket{{\cal R}BB}, \Braket{hEE}$ and $\Braket{hBB}$. Then, we found that $\Braket{{\cal R}EE}$ and $\Braket{hEE}$ ($\Braket{{\cal R}BB}$ and $\Braket{hBB}$) dominate $\Braket{{\cal R}BB}$ and $\Braket{hBB}$ ($\Braket{{\cal R}EE}$ and $\Braket{hEE}$) if $n = -2$ ($n = 2$). The contributions of electric and magnetic parts seem to turn over depending on the sign of $n$. We also confirmed that like the magnetic-scalar case \cite{Motta:2012rn}, the cross-bispectra include the term, which expresses the logarithmic growth as $\ln[-(k_1 + k_2 + k_3) \eta_I]$ and produces significant signals. 

The anisotropic stress, which consists of the square of the electromagnetic parts of the vector field, acts as a source of the CMB anisotropy. Hence, $\Braket{{\cal R}EE}, \Braket{{\cal R}BB}, \Braket{hEE}$ and $\Braket{hBB}$ induce not CMB bispectra but CMB power spectra. 
From formulation of such CMB power spectra, it was confirmed that they do not have the mode-coupling components, e.g., the scalar-vector correlation, and are independent of the coupling constant in the kinetic term of the vector field (\ref{eq:EM_action}).  

In numerical analysis for the case where the magnetic part is maximized, we observed that $C_\ell$'s generated from the electromagnetic-scalar and electromagnetic-tensor bispectra have similar shapes to those induced by the auto-correlations of the primary non-electromagnetic scalar and tensor perturbations, respectively. With respect to the amplitudes, the electromagnetic-scalar spectra are about $1\%$ of the scalar auto-correlated spectra. Interestingly, the former signals have opposite signs of the latter ones. In the tensor modes, although the signs are not flipped, the ratio between the electromagnetic-tensor and tensor auto-correlated spectra is improved to more than $10\%$. Especially, at small scales, the signals are enhanced and therefore the shape of the ${\cal B}{\cal B}$ mode changes compared with the primary non-electromagnetic case. The above characteristic behaviors may inform us of the existence of the coupling between the scalar and vector fields (\ref{eq:EM_action}). 

In this paper, we focused on the case where CMB signals are maximized, i.e., $n = 2.1$. However, their scale dependence and magnitudes are sensitive to $n$. This value directly reflects the model of megnetogenesis; hence detailed probes with actual observational data remain as a future issue. 

\acknowledgments
We would like to thank to Kiyotomo Ichiki for useful discussion. This work was supported in part by a Grant-in-Aid for JSPS Research under Grant Nos.~22-7477 (MS) and 24-2775 (SY), and Grant-in-Aid for Nagoya University Global COE Program ``Quest for Fundamental Principles in the Universe: from Particles to the Solar System and the Cosmos'', from the Ministry of Education, Culture, Sports, Science and Technology of Japan. 
We also acknowledge the Kobayashi-Maskawa Institute for the Origin of Particles and the Universe, Nagoya University, for providing computing resources useful in conducting the research reported in this paper. 

\appendix
\section{Projection vectors and tensors}\label{appen:polarization}

Here, let us summarize the conventions of the projection tensors based on refs.~\cite{Shiraishi:2010kd, Shiraishi:2012rm, Shiraishi:2012sn, Shiraishi:2012bh}. 
To calculate actual CMB power spectra, we shall fix the definition of a unit vector, a normalized divergenceless vector and a transverse-traceless tensor. These are respectively given by 
\begin{eqnarray}
\begin{split}
\hat{k}_a &= \sum_m \alpha_a^{m} Y_{1 m}(\hat{\bf k}) ~, \\
\epsilon_a^{(\pm 1)} (\hat{\bf k}) 
&= \mp \sum_m \alpha_a^m {}_{\pm 1} Y_{1 m} (\hat{\bf k})~, \\
e_{ab}^{(\pm 2)} (\hat{\bf k}) 
&=\sqrt{2} \epsilon_a^{(\pm 1)} (\hat{\bf k}) \epsilon_b^{(\pm 1)} (\hat{\bf k}) ~,
\end{split}
\end{eqnarray}
with
\begin{eqnarray}
\alpha_a^m \equiv \sqrt{\frac{2 \pi}{3}}
 \left(
  \begin{array}{ccc}
   -m (\delta_{m,1} + \delta_{m,-1}) \\
   i~ (\delta_{m,1} + \delta_{m,-1}) \\
   \sqrt{2} \delta_{m,0}
  \end{array}
\right)~.
\end{eqnarray}
Here, the contraction of $\alpha_a^m$'s is easily calculated as 
\begin{eqnarray}
\alpha_a^m \alpha_a^{m'} = \frac{4 \pi}{3} (-1)^m \delta_{m,-m'}~, \ \
\alpha_a^m \alpha_a^{m' *} = \frac{4 \pi}{3} \delta_{m,m'}~.
\end{eqnarray}
A divergenceless vector and a transverse-traceless tensor obey 
\begin{eqnarray}
\begin{split}
\hat{k}^a \epsilon_a^{(\pm 1)}(\hat{\bf k}) &= 0~, \\
\epsilon^{(\pm 1) *}_a (\hat{\bf k}) &= \epsilon^{(\mp 1)}_a (\hat{\bf k})
 = \epsilon^{(\pm 1)}_a (-\hat{\bf k})~, \\
\epsilon^{(\lambda)}_a (\hat{\bf k}) \epsilon^{(\lambda')}_a (\hat{\bf k}) 
&= \delta_{\lambda, -\lambda'} \ \ \ ({\rm for} \ \lambda, \lambda' = \pm 1) ~, \\
e_{aa}^{(\pm 2)}(\hat{\bf k}) &= \hat{k}_a e_{ab}^{(\pm 2)}(\hat{\bf k}) = 0~, \\
e_{ab}^{(\pm 2) *}(\hat{\bf k}) &= e_{ab}^{(\mp 2)}(\hat{\bf k}) = e_{ab}^{(\pm
2)}(- \hat{\bf k})~, \\
e_{ab}^{(\lambda)}(\hat{\bf k}) e_{ab}^{(\lambda')}(\hat{\bf k}) &= 2
\delta_{\lambda, -\lambda'} \ \ \ ({\rm for} \ \lambda, \lambda' = \pm 2)~. \label{eq:pol_tens_relation}
\end{split}
\end{eqnarray}
By a divergenceless vector and an antisymmetric tensor, a unit vector is also expressed as
\begin{eqnarray}
\hat{k}_c = i \eta_{abc} \epsilon_a^{(+1)}(\hat{\bf k}) 
\epsilon_b^{(-1)}(\hat{\bf k}) ~, \label{eq:P_eta_expand}
\end{eqnarray}
which is used in calculation of equation~(\ref{eq:contraction}). 

Obeying these notations, we define the projection vectors and tensors of the scalar, vector and tensor modes, respectively, as 
\begin{eqnarray}
\begin{split}
O_a^{(0)}(\hat{\bf k}) &\equiv i \hat{k}_a ~, \\
O_a^{(\pm 1)}(\hat{\bf k}) 
&\equiv - i \epsilon_a^{(\pm 1)}(\hat{\bf k})
~, 
\end{split} 
\end{eqnarray}
and 
\begin{eqnarray}
\begin{split}
O_{ab}^{(0)}(\hat{\bf k}) &\equiv - \hat{k}_a \hat{k}_b + \frac{1}{3} \delta_{ab} \\
&= - 2 I_{2 1 1}^{0 1 -1} \sum_{M m_a m_b} 
Y_{2 M}^*(\hat{\bf k}) \alpha_a^{m_a} \alpha_b^{m_b} 
\left(
  \begin{array}{ccc}
  2 & 1 & 1 \\
  M & m_a & m_b 
  \end{array}
 \right) ~, \\
O_{ab}^{(\pm 1)}(\hat{\bf k}) 
&\equiv 
\hat{k}_a {\epsilon}_b^{(\pm 1)}(\hat{\bf k}) 
+ \hat{k}_b {\epsilon}_a^{(\pm 1)}(\hat{\bf k}) \\
&= \pm 2\sqrt{3} I_{2 1 1}^{0 1 -1} \sum_{M m_a m_b} 
{}_{\mp 1} Y^*_{2 M}(\hat{\bf k}) \alpha_a^{m_a} \alpha_b^{m_b} 
\left(
  \begin{array}{ccc}
  2 & 1 & 1 \\
  M & m_a & m_b
  \end{array}
 \right)
~, \\
O_{ab}^{(\pm 2)}(\hat{\bf k}) &\equiv 
e_{ab}^{(\pm 2)}(\hat{\bf k})
 \\
&= 2\sqrt{3} I_{2 1 1}^{0 1 -1}
\sum_{M m_a m_b} {}_{\mp 2}Y_{2 M}^*(\hat{\bf k})
\alpha_a^{m_a} \alpha_b^{m_b}
 \left(
  \begin{array}{ccc}
  2 & 1 & 1 \\
  M & m_a & m_b
  \end{array}
 \right) ~, 
\end{split} 
\end{eqnarray}
where $I_{2 1 1}^{0 1 -1} = \sqrt{\frac{3}{8\pi}}$ is given by equation~(\ref{eq:I_sym}). These decompose arbitrary physical vector and tensor such as the velocity, the metric and the anisotropic stress into the scalar, vector and tensor components:
\begin{eqnarray}
\begin{split}
\omega_a({\bf k}) &=  \omega^{(0)}({\bf k}) O^{(0)}_a(\hat{\bf k}) 
+ \sum_{\lambda = \pm 1} \omega^{(\lambda)}({\bf k}) O^{(\lambda)}_a(\hat{\bf k})~, \\
\chi_{ab}({\bf k}) &= - \frac{1}{3} \chi_{\rm iso}({\bf k}) \delta_{ab} 
+ \chi^{(0)}({\bf k}) O^{(0)}_{ab}(\hat{\bf k}) \\
&\quad+ \sum_{\lambda = \pm 1} \chi^{(\lambda)}({\bf k}) O^{(\lambda)}_{ab}(\hat{\bf k})
+ \sum_{\lambda = \pm 2} \chi^{(\lambda)}({\bf k}) O^{(\lambda)}_{ab}(\hat{\bf k})
~. 
\end{split}
\end{eqnarray}
From equation~(\ref{eq:pol_tens_relation}), we have the inverse formulae as 
\begin{eqnarray}
\begin{split}
  \omega^{(0)}({\bf k}) &= - O_a^{(0)}({\bf k}) \omega_a({\bf k}) ~, \\
\omega^{(\pm 1)}({\bf k}) &= - O_a^{(\mp 1)}({\bf k}) \omega_a({\bf k}) ~, \\
\chi^{(0)}({\bf k}) &= \frac{3}{2}O^{(0)}_{ab}(\hat{\bf k}) \chi_{ab}({\bf k}) ~, \\
\chi^{(\pm 1)}({\bf k}) &= \frac{1}{2} O^{(\mp 1)}_{ab}(\hat{\bf k}) \chi_{ab}({\bf k})~, \\ 
\chi^{(\pm 2)}({\bf k}) &= \frac{1}{2} O_{ab}^{(\mp 2)}(\hat{\bf k}) \chi_{ab}({\bf k})~. \label{eq:tensor_inverse}
\end{split}
\end{eqnarray} 


\section{Time-dependent mode function}\label{appen:u}

Here, we explain properties of the time-dependent part of the mode functions, $u_{m}(x)$, given by equation (\ref{eq:u_func}). 

For specific $m$, we have the analytic formulae as
\begin{eqnarray}
u_{5/2}(x) &=& - \frac{1}{3} (-3 + 3ix + x^2) e^{ix} ~, \\
u_{3/2}(x) &=& (1 - ix) e^{ix} ~, \\
u_{-3/2}(x) &=& - \frac{3}{x^3} (i + x) e^{ix} ~.
\end{eqnarray}
The derivative with respect to $x$ is given by the recurrence formula as
\begin{eqnarray}
\frac{d u_m(x)}{dx} = \frac{2m}{x} 
\left[ u_m (x) - u_{m+1} (x) \right] ~. 
\end{eqnarray}
In particular, we have 
\begin{eqnarray}
\frac{d}{dx} u_{5/2}(x) 
&=& - \frac{1}{3} x \left( -1 + i x \right) e^{ix} ~, \\ 
\frac{d}{dx} u_{-3/2}(x) &=& 
3 \left( \frac{3i}{x^4} + \frac{3}{x^3} - \frac{i}{x^2} \right) e^{ix} ~.
\end{eqnarray}
At the limit: $x \to 0$, these asymptotically behave as 
\begin{eqnarray}
u_m(x \to 0) &=& 
\begin{cases}
1 & (m > 0) \\ 
0 & (m = 0) \\ 
\frac{4^{-m} \pi (i - \cot m \pi)}{\Gamma(m) \Gamma(m+1)} x^{2m} & (m < 0) 
\end{cases} ~, \label{eq:mode_lim} \\
\left(\frac{d u_m(x)}{d x}\right)_{x \to 0} &=& 
\begin{cases}
\frac{1}{2(m-1)} x & (m > \frac{1}{2}) \\
\frac{2^{1 - 2m} \pi (i - \cot m \pi)}{ \Gamma(m)^2 } x^{2m-1} & (m \leq \frac{1}{2}) 
\end{cases}
~. \label{eq:mode_dif_lim}
\end{eqnarray} 

\section{Specific expressions of $K^{{\cal R} YY}$ and $K^{h YY}$}\label{appen:integral}

\subsection{$\nu = \mu = 3/2$ and $\alpha = 5/2$}

The time-integlas in equations~(\ref{eq:time_int_scal}) and (\ref{eq:time_int_tens}) are  
\begin{eqnarray}
&& \int^{\infty}_{-k_1 \eta}d\tau_{1} \tau_{1}^{-4} 
u_{3/2}(\tau_{1}) 
\left( \frac{d}{d\tau_{1}}u_{5/2}(x_{2}\tau_{1}) \right)
\left( \frac{d}{d\tau_{1}}u_{5/2}(x_{3}\tau_{1}) \right) 
\nonumber \\ 
&&\qquad   
= \frac{k_2^2 k_3^2}{9 (-k_1 \eta) k_1^4} 
+ i J_{5/2}^{(1a)} + {\cal O}(-k_1 \eta) 
~, \\ 
&& \int^{\infty}_{-k_1 \eta}d\tau_{1} \tau_{1}^{-4} 
\left( \tau_1 \frac{d}{d\tau_1} u_{3/2}(\tau_{1}) \right)
\left( \frac{d}{d\tau_{1}}u_{5/2}(x_{2}\tau_{1}) \right)
\left( \frac{d}{d\tau_{1}}u_{5/2}(x_{3}\tau_{1}) \right) 
\nonumber \\ 
&&\qquad  
= i J_{5/2}^{(1b)} + {\cal O}(-k_1 \eta) 
~,\\
&& \int^{\infty}_{-k_1 \eta} d\tau_{1} \tau^{-4}_{1} 
 u_{3/2}(\tau_{1}) 
u_{5/2}(x_{2}\tau_{1}) 
u_{5/2}(x_{3}\tau_{1}) 
\nonumber \\ 
&&\qquad   
= \frac{1}{3 (-k_1 \eta)^3} + \frac{3k_1^2 + k_2^2 + k_3^2}{6 (-k_1 \eta) k_1^2}
+ \frac{\pi}{6} + i J_{5/2}^{(2a)}  + {\cal O}(-k_1 \eta) ~,  
\\ 
&& \int^{\infty}_{-k_1 \eta} d\tau_{1} \tau^{-4}_{1} 
\left( \tau_{1}\frac{d}{d\tau_{1}} u_{3/2}(\tau_{1}) \right)
u_{5/2}(x_{2}\tau_{1}) 
u_{5/2}(x_{3}\tau_{1})
\nonumber \\ 
&&\qquad   
= \frac{1}{(-k_1 \eta)} + \frac{\pi}{2} 
+ i J_{5/2}^{(2b)}  + {\cal O}(-k_1 \eta) ~,  
\\ 
&& \int^{\infty}_{-k_1 \eta} 
d\tau_{1} \tau^{-4}_{1} u_{3/2}(\tau_{1}) 
u_{5/2}(x_{2}\tau_{1}) 
\left( \tau_{1}\frac{d}{d\tau_{1}}u_{5/2}(x_{3}\tau_{1}) \right) 
\nonumber \\ 
&&\qquad   
= \frac{k_3^2}{3 (-k_1 \eta) k_1^2} 
+ i J_{5/2}^{(3a)} + {\cal O}(-k_1 \eta)
 ~, \\
&& \int^{\infty}_{-k_1 \eta} 
d\tau_{1} \tau^{-4}_{1} u_{3/2}(\tau_{1}) 
\left( \tau_{1}\frac{d}{d\tau_{1}}
u_{5/2}(x_{2}\tau_{1}) \right)
u_{5/2}(x_{3}\tau_{1}) 
\nonumber \\ 
&&\qquad   
= \frac{k_2^2}{3 (-k_1 \eta) k_1^2} + i  J_{5/2}^{(3b)} + {\cal O}(-k_1 \eta)~,
\end{eqnarray}
where 
\begin{eqnarray}
J_{5/2}^{(1a)} &=& \frac{ k_2^2 k_3^2}{9 k_1^5 \omega^2} 
\left[ \omega^3 + (k_2+k_3)(k_2 k_3- \omega^2) + \omega(k_2^2 + k_3^2)\right] ~, \\ 
J_{5/2}^{(1b)} &=& \frac{k_2^2 k_3^2}{9 k_1^3 \omega^3}
\left[  \omega^2 + \omega(k_2+k_3)+2 k_2 k_3 \right] ~, \\
J_{5/2}^{(2a)} &=& - \frac{1}{3}\left[\gamma + \ln(- \omega \eta) \right] 
\nonumber \\
&& + \frac{1}{9 k_1^3 \omega^2}
\left[ k_2^2 k_3^2(k_1+\omega) -3 \omega k_2 k_3(\omega^2 - k_1 \omega + k_1^2)+3 k_1 \omega^3(2 k_1 - \omega) 
+ \omega^5 \right] ~, \\ 
J_{5/2}^{(2b)} &=& 
-  \left[ \gamma + \ln(- \omega \eta) \right]
\nonumber \\
&& + \frac{1}{9 k_1 \omega^3} 
\left[ 9 \omega^4 - 3 \omega^2(k_2^2 + 4 k_2 k_3 + k_3^2)+ 3 \omega k_1 k_2 k_3 - 2 k_2^2 k_3^2\right] ~, \\ 
J_{5/2}^{(3a)} &=& \frac{ k_3^2}{9 k_1^3 \omega^3} 
\left[
3 \omega^4 -3 (k_2 + k_3)\omega^3 + (k_2^2 + 3 k_3^2)\omega^2 \right. \nonumber \\ 
&&\left. \qquad\quad + k_2(k_2^2 + k_2 k_3 + 3 k_3^2)\omega + 2 k_2^2 k_3(k_2+k_3)\right] ~, \\
J_{5/2}^{(3b)} &=& J_{5/2}^{(3a)}(k_2 \leftrightarrow k_3) ~. 
\end{eqnarray} 
The other parts are expanded as  
\begin{eqnarray}
&& u_{3/2}^*(-k_1 \eta) u_{5/2}^*(-k_2 \eta) u_{5/2}^*(-k_3 \eta) 
\nonumber \\ 
&&\qquad
= 1 + \frac{1}{6} 
\left(3 k_1^2 + k_2^2 + k_3^2 \right) \eta^2 - \frac{i}{3} (- k_1 \eta)^3 + {\cal O}((-k_1 \eta)^4) ~, \\
&& u_{3/2}^*(-k_1 \eta) u_{5/2}'^*(-k_2 \eta) u_{5/2}'^*(-k_3 \eta) 
\nonumber \\ 
&&\qquad 
= \frac{k_2^2 k_3^2 \eta^2}{9} 
\left[ 1 + \frac{1}{2} (k_1^2 + k_2^2 + k_3^2) \eta^2 
+ \frac{i}{3} (k_1^3 + k_2^3 + k_3^3) \eta^3 \right] + {\cal O}((-k_1 \eta)^6) ~.
\end{eqnarray}
Picking up the leading-order terms in the combinations of these equations, we can reach the final expressions of $K^{{\cal R}YY}$'s and $K^{hYY}$'s  at the end of inflation on the superhorizon scales ($-k \eta_I \ll 1$):
\begin{eqnarray}
K_{1}^{{\cal R}EE}(\alpha = 5/2) 
&=& - \frac{4}{9} k_1 k_2^2 k_3^2 \eta_I^2 
\left( J_{5/2}^{(1a)} + J_{5/2}^{(1b)} \right) ~, \\ 
K_{2}^{{\cal R}EE}(\alpha = 5/2) 
&=&  \frac{2 k_2^3 k_3^3 \eta_I^2}{9k_1}  
\left( J_{5/2}^{(2a)} + J_{5/2}^{(2b)} - \frac{k_1^3 + k_2^3 + k_3^3}{9 k_1^3} \right) ~, \\ 
K_{3}^{{\cal R}EE}(\alpha = 5/2) 
&=& 
\frac{2 k_2^3 k_3^3 \eta_I^2}{9 k_1} 
J_{5/2}^{(3a)} ~, \\ 
K_{4}^{{\cal R}EE}(\alpha = 5/2) 
&=& \frac{2 k_2^3 k_3^3 \eta_I^2}{9 k_1}  J_{5/2}^{(3b)} ~, \\
K_{1}^{{\cal R}BB}(\alpha = 5/2) 
&=& - 4k_{1}k_{2}k_{3} 
\left( J_{5/2}^{(1a)} + J_{5/2}^{(1b)} \right) ~, \\ 
K_{2}^{{\cal R}BB}(\alpha = 5/2) 
&=& 2\frac{k^{2}_{2}k^{2}_{3}}{k_{1}} 
\left( J_{5/2}^{(2a)} + J_{5/2}^{(2b)} -\frac{1}{9} \right) ~, \\ 
K_{3}^{{\cal R}BB}(\alpha = 5/2) 
&=& 2\frac{k^{2}_{2}k^{2}_{3}}{k_{1}} J_{5/2}^{(3a)} ~, \\ 
K_{4}^{{\cal R}BB}(\alpha = 5/2) 
&=& 2\frac{k^{2}_{2}k^{2}_{3}}{k_{1}} J_{5/2}^{(3b)} ~, 
\end{eqnarray}
and 
\begin{eqnarray}
K_{1}^{hEE}(\alpha = 5/2) 
&=& \frac{2}{9} k_{1} k_2^2 k_3^2 \eta_I^2 J_{5/2}^{(1a)} ~, \\ 
K_{2}^{hEE}(\alpha = 5/2) 
&=&  \frac{2 k_2^3 k_3^3 \eta_I^2}{9k_1}  
\left( J_{5/2}^{(2a)} - \frac{k_1^3 + k_2^3 + k_3^3}{9 k_1^3} \right) ~, \\ 
K_{1}^{hBB}(\alpha = 5/2) 
&=& 2 k_{1}k_{2}k_{3} J_{5/2}^{(1a)}  ~, \\ 
K_{2}^{hBB}(\alpha = 5/2) 
&=& 2\frac{k^{2}_{2}k^{2}_{3}}{k_{1}} 
\left( J_{5/2}^{(2a)} -\frac{1}{9} \right) ~.
\end{eqnarray}

\subsection{$\nu = \mu = 3/2$ and $\alpha = -3/2$}

The time integrals and the product of the mode functions are resectively expanded as
\begin{eqnarray}
&& \int^{\infty}_{-k_1 \eta}d\tau_{1} \tau_{1}^{4} 
u_{3/2}(\tau_{1}) 
\left( \frac{d}{d\tau_{1}}u_{-3/2}(x_{2}\tau_{1}) \right)
\left( \frac{d}{d\tau_{1}}u_{-3/2}(x_{3}\tau_{1}) \right) 
\nonumber \\ 
&&\qquad 
= - \frac{27 k_1^6}{k_2^3 k_3^3 (-k_1 \eta)^3} 
- \frac{27 k_1^4 (3k_1^2 + k_2^2 + k_3^2)}{2 k_2^3 k_3^3 (-k_1 \eta)}
- \frac{27 \pi k_1^6}{2 k_2^3 k_3^3}
+ i J_{-3/2}^{(1a)} + {\cal O}(-k_1 \eta) 
~, \\ 
&& \int^{\infty}_{-k_1 \eta}d\tau_{1} \tau_{1}^{4} 
\left( \tau_1 \frac{d}{d\tau_1} u_{3/2}(\tau_{1}) \right)
\left( \frac{d}{d\tau_{1}}u_{-3/2}(x_{2}\tau_{1}) \right)
\left( \frac{d}{d\tau_{1}}u_{-3/2}(x_{3}\tau_{1}) \right) 
\nonumber \\ 
&&\qquad 
= - \frac{81 k_1^6}{k_2^3 k_3^3 (-k_1 \eta)} - \frac{81 \pi k_1^6}{2 k_2^3 k_3^3}+   
i J_{-3/2}^{(1b)} + {\cal O}(-k_1 \eta) 
~,\\
&& \int^{\infty}_{-k_1 \eta} d\tau_{1} \tau^{4}_{1} 
 u_{3/2}(\tau_{1}) 
u_{-3/2}(x_{2}\tau_{1}) 
u_{-3/2}(x_{3}\tau_{1}) 
\nonumber \\ 
&&\qquad 
= - \frac{9 k_1^6}{k_2^3 k_3^3 (-k_1 \eta)} + i J_{-3/2}^{(2a)}  + {\cal O}(-k_1 \eta) ~,  
\\ 
&& \int^{\infty}_{-k_1 \eta} d\tau_{1} \tau^{4}_{1} 
\left( \tau_{1}\frac{d}{d\tau_{1}} u_{3/2}(\tau_{1}) \right)
u_{-3/2}(x_{2}\tau_{1}) 
u_{-3/2}(x_{3}\tau_{1}) 
= i J_{-3/2}^{(2b)}  + {\cal O}(-k_1 \eta) ~,  
\\ 
&& \int^{\infty}_{-k_1 \eta} 
d\tau_{1} \tau^{4}_{1} u_{3/2}(\tau_{1}) 
u_{-3/2}(x_{2}\tau_{1}) 
\left( \tau_{1}\frac{d}{d\tau_{1}}u_{-3/2}(x_{3}\tau_{1}) \right) 
\nonumber \\
&&\qquad  
= \frac{27 k_1^6}{k_2^3 k_3^3 (-k_1 \eta)} 
+ i J_{-3/2}^{(3a)} + {\cal O}(-k_1 \eta)
 ~, \\
&& \int^{\infty}_{-k_1 \eta} 
d\tau_{1} \tau^{4}_{1} u_{3/2}(\tau_{1}) 
\left( \tau_{1}\frac{d}{d\tau_{1}}
u_{-3/2}(x_{2}\tau_{1}) \right)
u_{-3/2}(x_{3}\tau_{1}) 
\nonumber \\
&&\qquad  
= \frac{27 k_1^6}{k_2^3 k_3^3 (-k_1 \eta)}  + i  J_{-3/2}^{(3b)} + {\cal O}(-k_1 \eta) ~,
\end{eqnarray}
with
\begin{eqnarray}
J_{-3/2}^{(1a)} &=& \frac{9 k_1^6}{k_2^3 k_3^3} 
\left[ 3 [\gamma + \ln(- \omega \eta)] - 4 \right. \nonumber \\ 
&&\left. \qquad
- \frac{1}{k_1^3} 
\left\{ (3 k_1^2 + k_2^2) k_3 + k_3^3 + \frac{k_1 k_2^2}{\omega^2} (k_1 + k_2)^2 + \frac{k_2}{\omega} (3 k_1^3 + 2 k_1^2 k_2 + k_2^3 ) \right\}
\right]
~, \nonumber \\
\\
J_{-3/2}^{(1b)} &=&  \frac{ 9 k_1^6}{k_2^3 k_3^3} 
\left[  
9 [ \gamma + \ln(- \omega \eta) ]
- \frac{9 \omega}{k_1}  \right. \nonumber \\ 
&&\left. \qquad\quad
+ \frac{1}{k_1 \omega^3} 
\{ 3 k_2^2 (k_1 + k_2)^2 + 
  9 k_2 (k_1 + k_2) (k_1 + 2 k_2) k_3 \right. \nonumber \\ 
&&\left.\qquad\qquad\qquad
+ (3 k_1^2 + 27 k_1 k_2 + 32 k_2^2) k_3^2 + 6 (k_1 + 3 k_2) k_3^3 + 3 k_3^4 \}
\right]
~, \\ 
J_{-3/2}^{(2a)} &=& 
- \frac{9 k_1^6}{k_2^3 k_3^3} 
\left[ \frac{k_3}{k_1} + \frac{k_2}{\omega^2} (k_1 + k_2) 
+ \frac{k_1^2 + k_2^2}{k_1 \omega} \right] ~, \\
J_{-3/2}^{(2b)} &=& 
-\frac{9 k_1^7}{k_2^3 k_3^3 \omega^3} 
[k_1^2 + 3 k_1 (k_2 + k_3) + 2 (k_2^2 + 3 k_2 k_3 + k_3^2)]
 ~, \\ 
J_{-3/2}^{(3a)} &=& \frac{9 k_1^6 }{k_2^3 k_3^3} 
\left[ 
\frac{3 k_2}{k_1} + \frac{2 k_3^2}{\omega^3} (k_1 + k_3) 
+ \frac{k_3}{k_1 \omega^2} (3 k_1^2 + k_1 k_3 + k_3^2) 
+ \frac{3 k_1^2 + k_3^2}{ k_1 \omega} \right] ~, \\ 
J_{-3/2}^{(3b)} &=& J_{-3/2}^{(3a)}(k_2 \leftrightarrow k_3) ~,
\end{eqnarray}
and 
\begin{eqnarray}
&& u_{3/2}^*(-k_1 \eta) u_{-3/2}^*(-k_2 \eta) u_{-3/2}^*(-k_3 \eta) \nonumber \\  
&&\qquad = - \frac{9}{k_2^3 k_3^3 \eta^6} 
\left[ 1 + \frac{1}{2}(k_1^2 + k_2^2 + k_3^2) \eta^2 
+ \frac{i}{3} (k_1^3 + k_2^3 + k_3^3) \eta^3 
 \right] + {\cal O}((-k_1 \eta)^{-2}) ~, \\
&& u_{3/2}^*(-k_1 \eta) u_{-3/2}'^*(-k_2 \eta) u_{-3/2}'^*(-k_3 \eta) \nonumber \\  
&&\qquad
= - \frac{81}{k_2^3 k_3^3 \eta^8 } 
\left[ 1 + \frac{1}{6} (3 k_1^2 + k_2^2 + k_3^2) \eta^2
- \frac{i}{3} (- k_1 \eta)^3  
\right]
+ {\cal O}((-k_1 \eta)^{-4}) ~.
\end{eqnarray}
In the same manner as the above case, we can gain
\begin{eqnarray}
K_{1}^{{\cal R}EE}(\alpha = -3/2) 
&=& \frac{324 k_1}{k_2^3 k_3^3 \eta_I^8} 
\left[ J_{-3/2}^{(1a)} + J_{-3/2}^{(1b)} + \frac{9 k_1^6}{k_2^3 k_3^3} \right] ~, \\ 
K_{2}^{{\cal R}EE}(\alpha = -3/2) 
&=& - \frac{162}{k_1 k_2^2 k_3^2 \eta_I^8}  
\left( J_{-3/2}^{(2a)} + J_{-3/2}^{(2b)} \right) ~, \\ 
K_{3}^{{\cal R}EE}(\alpha = -3/2) 
&=& - \frac{162}{k_1 k_2^2 k_3^2 \eta_I^8}   J_{-3/2}^{(3a)} ~, \\ 
K_{4}^{{\cal R}EE}(\alpha = -3/2) 
&=& - \frac{162}{k_1 k_2^2 k_3^2 \eta_I^8} J_{-3/2}^{(3b)} ~, \\ 
K_{1}^{{\cal R}BB}(\alpha = -3/2) 
&=& \frac{36 k_1}{k_2^2 k_3^2 \eta_I^6}  
\left[ J_{-3/2}^{(1a)} + J_{-3/2}^{(1b)} + \frac{9 k_1^3}{k_2^3 k_3^3} (k_1^3 + k_2^3 + k_3^3) \right] ~, \\ 
K_{2}^{{\cal R}BB}(\alpha = -3/2) 
&=& - \frac{18}{k_1 k_2 k_3 \eta_I^6}  
\left( J_{-3/2}^{(2a)} + J_{-3/2}^{(2b)} \right) ~, \\ 
K_{3}^{{\cal R}BB}(\alpha = -3/2) 
&=& - \frac{18}{k_1 k_2 k_3 \eta_I^6} J_{-3/2}^{(3a)} ~, \\ 
K_{4}^{{\cal R}BB}(\alpha = -3/2) 
&=& - \frac{18}{k_1 k_2 k_3 \eta_I^6} J_{-3/2}^{(3b)} ~,
\end{eqnarray}
and 
\begin{eqnarray}
K_{1}^{hEE}(\alpha = -3/2) 
&=& - \frac{162 k_1}{k_2^3 k_3^3 \eta_I^8} 
\left( J_{-3/2}^{(1a)} + \frac{9 k_1^6}{k_2^3 k_3^3} \right) ~, \\ 
K_{2}^{hEE}(\alpha = -3/2) 
&=& - \frac{162}{k_1 k_2^2 k_3^2 \eta_I^8} J_{-3/2}^{(2a)}  ~, \\ 
K_{1}^{hBB}(\alpha = -3/2) 
&=& - \frac{18 k_1}{k_2^2 k_3^2 \eta_I^6}  
\left[ J_{-3/2}^{(1a)}  + \frac{9 k_1^3}{k_2^3 k_3^3} (k_1^3 + k_2^3 + k_3^3) \right] ~, \\ 
K_{2}^{hBB}(\alpha = -3/2) 
&=& - \frac{18}{k_1 k_2 k_3 \eta_I^6} J_{-3/2}^{(2a)}  ~.
\end{eqnarray}

\bibliography{paper}
\end{document}